\documentclass[twocolumn]{aastex63}
\usepackage{graphicx}

\usepackage{placeins}
\usepackage{amsmath}

\newcommand{\kms}{km s$^{-1}$}

\begin{document}

\title{Metallicity and $\alpha$-element Abundance Gradients along the Sagittarius Stream as Seen by APOGEE}


\author[0000-0003-2969-2445]{Christian R. Hayes}
\affiliation{Department of Astronomy, University of Virginia, Charlottesville, VA 22904-4325, USA}

\author{Steven R. Majewski}
\affiliation{Department of Astronomy, University of Virginia, Charlottesville, VA 22904-4325, USA}

\author{Sten Hasselquist}
\altaffiliation{NSF Astronomy and Astrophysics Postdoctoral Fellow}
\affiliation{Department of Physics \& Astronomy, University of Utah, Salt Lake City, UT, 84112, USA}

\author{Borja Anguiano}
\affiliation{Department of Astronomy, University of Virginia, Charlottesville, VA 22904-4325, USA}

\author{Matthew Shetrone}
\affiliation{University of Texas at Austin, McDonald Observatory, McDonald Observatory, TX 79734, USA}

\author{David R. Law}
\affiliation{Space Telescope Science Institute, 3700 San Martin Drive, Baltimore, MD 21218}

\author{Ricardo P. Schiavon}
\affiliation{Astrophysics Research Institute, Liverpool John Moores University, 146 Brownlow Hill, Liverpool L3 5RF, UK}

\author{Katia Cunha}
\affiliation{Observat\'{o}rio Nacional, 77 Rua General Jos\'{e} Cristino, Rio de Janeiro, 20921-400, Brazil}
\affiliation{Steward Observatory, University of Arizona, 933 North Cherry Avenue, Tucson, AZ 85721, USA}

\author{Verne V. Smith}
\affiliation{National Optical Astronomy Observatory, 950 North Cherry Avenue, Tucson, AZ 85719, USA}

\author[0000-0002-1691-8217]{Rachael L. Beaton}
\altaffiliation{Hubble Fellow}
\altaffiliation{Carnegie-Princeton Fellow}
\affiliation{Department of Astrophysical Sciences, Princeton University, 4 Ivy Lane, Princeton, NJ~08544}
\affiliation{The Observatories of the Carnegie Institution for Science, 813 Santa Barbara St., Pasadena, CA~91101}

\author{Adrian M. Price-Whelan}
\affiliation{Department of Astrophysical Sciences, Princeton University, 4 Ivy Lane, Princeton, NJ~08544}
\affiliation{Center for Computational Astrophysics, Flatiron Institute, 162 Fifth Ave., New York, NY 10010, USA}

\author{Carlos Allende Prieto}
\affiliation{Instituto de Astrof\'{i}sica de Canarias (IAC), V\'{i}a L\'{a}ctea, E-38205 La Laguna, Tenerife, Spain}
\affiliation{Departamento de Astrof\'isica, Universidad de La Laguna (ULL), E-38206 La Laguna, Tenerife, Spain}

\author{Giuseppina Battaglia}
\affiliation{Instituto de Astrof\'{i}sica de Canarias (IAC), V\'{i}a L\'{a}ctea, E-38205 La Laguna, Tenerife, Spain}
\affiliation{Departamento de Astrof\'isica, Universidad de La Laguna (ULL), E-38206 La Laguna, Tenerife, Spain}

\author{Dmitry Bizyaev}
\affiliation{Apache Point Observatory and New Mexico State University, P.O. Box 59, Sunspot, NM, 88349-0059, USA}
\affiliation{Sternberg Astronomical Institute, Moscow State University, Moscow, Russia}

\author{Joel R. Brownstein}
\affiliation{Department of Physics and Astronomy, University of Utah, 115 S. 1400 E., Salt Lake City, UT 84112, USA}

\author{Roger E. Cohen}
\affiliation{Space Telescope Science Institute, 3700 San Martin Drive, Baltimore, MD 21218, USA}


\author{Peter M. Frinchaboy}
\affiliation{Department of Physics \& Astronomy, Texas Christian University, Fort Worth, TX 76129, USA}

\author{D. A. Garc\'{i}a-Hern\'{a}ndez}
\affiliation{Instituto de Astrof\'{i}sica de Canarias (IAC), V\'{i}a L\'{a}ctea, E-38205 La Laguna, Tenerife, Spain}
\affiliation{Departamento de Astrof\'isica, Universidad de La Laguna (ULL), E-38206 La Laguna, Tenerife, Spain}

\author{Ivan Lacerna}
\affiliation{Instituto de Astronom\'ia y Ciencias Planetarias, Universidad de Atacama, Copayapu 485, Copiap\'o, Chile}
\affiliation{Instituto Milenio de Astrof\'isica, Av. Vicu\~na Mackenna 4860, Macul, Santiago, Chile}

\author{Richard R. Lane}
\affiliation{Millennium Institute of Astrophysics, Av. Vicu\~na Mackenna 4860, 782-0436 Macul, Santiago, Chile}

\author{Szabolcs M\'esz\'aros}
\affiliation{ELTE E\"otv\"os Lor\'and University, Gothard Astrophysical Observatory, 9700 Szombathely, Szent Imre h. st. 112, Hungary}
\affiliation{Premium Postdoctoral Fellow of the Hungarian Academy of Sciences}
\affiliation{MTA-ELTE Exoplanet Research Group, 9700 Szombathely, Szent Imre h. st. 112, Hungary}

\author{Christian Moni Bidin}
\affiliation{Instituto de Astronom\'ia, Universidad Cat\'olica del Norte, Av. Angamos 0610, Antofagasta, Chile}

\author{Ricardo R. M\~unoz}
\affiliation{Universidad de Chile, Av. Libertador Bernardo O'Higgins 1058, Santiago de Chile}

\author{David L. Nidever}
\affiliation{Department of Physics, Montana State University, P.O. Box 173840, Bozeman, MT 59717-3840}
\affiliation{National Optical Astronomy Observatory, 950 North Cherry Avenue, Tucson, AZ 85719, USA}

\author{Audrey Oravetz}
\affiliation{Apache Point Observatory and New Mexico State University, P.O. Box 59, Sunspot, NM, 88349-0059, USA}

\author{Daniel Oravetz}
\affiliation{Apache Point Observatory and New Mexico State University, P.O. Box 59, Sunspot, NM, 88349-0059, USA}

\author{Kaike Pan}
\affiliation{Apache Point Observatory and New Mexico State University, P.O. Box 59, Sunspot, NM, 88349-0059, USA}

\author{Alexandre Roman-Lopes}
\affiliation{Departamento de F\'isica, Facultad de Ciencias, Universidad de La Serena, Cisternas 1200, La Serena, Chile}

\author{Jennifer Sobeck}
\affiliation{Department of Astronomy, University of Washington, Box 351580, Seattle, WA 98195, USA}

\author{Guy Stringfellow}
\affiliation{Center for Astrophysics and Space Astronomy, Department of Astrophysical and Planetary Sciences, University of Colorado, 389 UCB, Boulder, CO 80309-0389, USA}

\email{crh7gs@virginia.edu}

\begin{abstract}

Using 3D positions and kinematics of stars relative to the Sagittarius (Sgr) orbital plane and angular momentum, we identify 166 Sgr stream members observed by the Apache Point Observatory Galactic Evolution Experiment (APOGEE) that also have {\it Gaia} DR2 astrometry.  This sample of 63/103 stars in the Sgr trailing/leading arm are combined with an APOGEE sample of { 710} members of the Sgr dwarf spheroidal core (385 of them newly presented here) to establish differences of 0.6 dex in median metallicity and 0.1 dex in [$\alpha$/Fe] between our Sgr core and dynamically older stream samples.  Mild chemical gradients are found internally along each arm, but these steepen when anchored by core stars.  With a model of Sgr tidal disruption providing estimated dynamical ages (i.e., stripping times) for each stream star, we find a mean metallicity gradient of $0.12 \pm 0.03$ dex/Gyr for stars stripped from Sgr over time.  For the first time, an [$\alpha$/Fe] gradient is also measured within the stream, at $0.02 \pm 0.01$ dex/Gyr using magnesium abundances and $0.04 \pm 0.01$ dex/Gyr using silicon, which imply that the Sgr progenitor had significant radial abundance gradients.  We discuss the magnitude of those inferred gradients and their implication for the nature of the Sgr progenitor within the context of the current family of Milky Way satellite galaxies, and suggest that more sophisticated Sgr models are needed to properly interpret the growing chemodynamical detail we have on the Sgr system.

\end{abstract}


\section{Introduction}

The Sagittarius (Sgr) dwarf spheroidal (dSph) galaxy and its tidal stream provide a nearby and vivid example of a tidally disrupting dwarf galaxy \citep{ibata1994,maj03} and the hierarchical growth of large galaxies through minor mergers.  Because Sgr is in a quite advanced stage of tidal stripping, yet its stars are not yet fully mixed with those of the Milky Way (MW), the system has become a remarkably versatile tool for exploring a great variety of astrophysical problems. 

Numerous studies have exploited the extensive tidal debris structure as a sensitive probe of the MW, its dark matter content, and its dynamics.  For example, because Sgr's tidal arms wrap through a large extent of the MW halo and trace the past and future orbit of the core, they can constrain the 3D shape of the MW's dark matter halo \citep{helmi2004,johnston2005,law2005,lm10,deg2013,ibata2013,veraciro2013}.  Moreover, the alignment of Sgr's orbit is nearly perpendicular to the MW disk and crosses the disk midplane relatively near the Sun-Galactic Center axis; this fortuitous configuration means that the solar rotational velocity can also be gauged directly via the reflex solar motion imprinted in the velocities/proper motions of stars in the stream \citep{majewski2006,lm10,carlin2012,hayes2018c}.  Sgr has also been identified as a possible culprit for dynamical perturbations observed in the MW disk, and as such provides a case study on the potential effects of minor mergers on the evolution of the stellar and HI disks \citep{ibata1998,gomez2013,lap18,laporte2019}.  

Obviously, the Sgr system also lends uniquely accessible and detailed insights into the tidal disruption and dynamical evolution of dwarf galaxy satellites.  This includes not only clues into potential morphological and dynamical changes in dwarf galaxies induced by the encounters with larger galaxies like the MW \citep{lokas2010,penarrubia2010,penarrubia2011,frinchaboy2012,lokas2012,majewski2013}, but also effects on their star formation histories and the chemical evolution of their stellar populations.  The latter have clearly been shaped by the interplay between episodic star formation incited by gravitational shocking at orbital pericenter and the stripping of gas \citep{siegel2007,teppergarcia2018}.

A particularly important lesson learned from studies of the Sgr system is that any assessment of the chemical and star formation histories and distribution functions of Sgr or another tidally disrupted system will be incomplete and biased without properly accounting for the stellar populations lost via tidal stripping \citep[][; see also earlier discussions of this phenomenon in \citealt{majewski2002,munoz2006}]{chou2010,carlin2018}.  This is because tidal stripping preferentially acts on the least bound stars in a dwarf, and those stars tend to be older and less chemically evolved stars in the system.

The discovery of large mean metallicity differences at the level of $\Delta$[Fe/H] $\sim$ 0.4-0.6 dex between samples of stars in the Sgr core and the (lower metallicity) Sgr stream \citep{chou2007,monaco2007} provided early suggestions of the possible metallicity gradients along the Sgr stream.  If such chemical gradients do indeed exist along the Sgr stream, they may be the preserved remnants of chemical gradients that existed within the Sgr progenitor galaxy.

\defcitealias{lm10}{LM10}

N-body modeling of Sgr's tidal stripping was implemented by \citet[][hereafter LM10]{lm10}, who used a prescription for assigning metallicities to model particles based on their initial energy in the bound progenitor, which naturally yielded a radial gradient in mean metallicity in the simulated dwarf.  Based on this modeling, \citetalias{lm10} found that the observed metallicity differences between stream and core implied a mean radial metallicity variation as large as 2.0 dex before Sgr's tidal disruption, exceeding that seen in any other dwarf galaxy.

Since these first identifications of significant metallicity differences between the Sgr stream and core, further studies have measured the metallicity of Sgr stream stars and reported metallicity gradients \citep{keller2010,carlin2012,shi2012,hyde2015} along the Sgr stream.  However, the sampling and measurement of gradients was not consistent across studies, which complicates their comparison.  Specifically, some authors report metallicity gradients from the Sgr core through each tidal arm (such that the high end of the gradient is anchored by the metallicity of the core) and find metallicity gradients of about 2.4-2.7 $\times 10^{-3}$ dex deg$^{-1}$ along the trailing arm \citep{keller2010,hyde2015}.   Other studies measure only the internal metallicity gradients within each arm (excluding the metallicity of the Sgr dSph core), which produces much flatter gradients, around 1.4-1.8 $\times$ $10^{-3}$ dex deg$^{-1}$ along the trailing arm and about 1.5 $\times$ $10^{-3}$ dex deg$^{-1}$ along the leading arm \citep{carlin2012,shi2012}. 

Because a large fraction ($75\%$ at high latitudes) of the MW halo M giants belong to the Sgr stream \citep{maj03}, some studies have employed a color selection to exclusively study the relatively metal-rich M giants in the stream, since they are subject to less contamination than samples of the more common K giants \citep{chou2007,monaco2007,keller2010,carlin2018}.  However, M giants are only produced by higher metallicity populations, so these samples would have an implicit metallicity bias, and could skew some of these past measurements of metallicity gradients.

Because of the observational demands required by high-resolution spectroscopy, few detailed chemical abundance studies of stream stars have been performed, and only measured abundances for relatively small samples \citep{monaco2007,chou2010,keller2010,carlin2018}.  However, such studies have attempted to explore the $\alpha$-element abundances of Sgr stream stars, and typically report similar $\alpha$-element abundance levels to stars in the Sgr core \citet{monaco2007,chou2010,carlin2018}, or equivalently suggest no significant $\alpha$-element gradients along the stream \citep{keller2010}. 

The Apache Point Observatory Galactic Evolution Experiment \citep[APOGEE;][]{apogee} provides a unique opportunity to study the detailed chemistry of the Sgr stream.  APOGEE is a high-resolution ($R \sim 22,500$), $H$-band (1.5-1.7 $\mu$m) spectroscopic survey that primarily targets red giant stars and samples a relatively large area of the sky.  While the APOGEE survey imposes a blue color limit to prioritize observations of red giants and minimize contamination from warmer main sequence dwarfs, this limit, $(J-K)_0 \geq 0.3$ in halo fields ($|b| \gtrsim 16^{\circ}$, where most of the Sgr stream lies) and $(J-K)_0 \geq 0.5$ otherwise, is still liberal enough to provide relatively unbiased metallicity coverage for red giant branch stars \citep[RGB;][]{zas13,zas17}.  In addition, the dual-hemisphere coverage of APOGEE-2 allows us to sample nearly continuously along large sections of both arms of the Sgr stream.

Observations of Sgr dSph core members were first reported in APOGEE by \citet{majewski2013} using the Sloan Digital Sky Survey (SDSS) Data Release 12 \citep[DR12][]{dr12}, and the membership was expanded by \citet{has17} using SDSS DR13 \citep{dr13,holtzman2018}, both taking advantage of the intentional APOGEE targeting of the Sgr core.  While a few APOGEE fields were placed intentionally along the Sgr stream,
\citet{hasselquist2019} demonstrated that both the trailing and leading arm of the Sgr stream are relatively well-sampled {\it serendipitously} by the random targeting employed by the APOGEE survey.  

\citet{hasselquist2019} used chemical tagging to identify 35 relatively metal-rich, [Fe/H] $\gtrsim -1.2$, Sgr stream stars in the APOGEE data presented in SDSS DR14 \citep{dr14,holtzman2018}, which only included APOGEE data in the Northern Hemisphere.  However, the chemical tagging method that was used to identify these Sgr stars is limited to these higher metallicities, because it relies on the fact that the chemical abundance profile of Sgr is distinct from the MW at these metallicities \citep{has17,hasselquist2019}.  

At lower metallicities, the chemical abundance profile of Sgr begins to merge with that of the accreted MW halo \citep{hayes2018a,hasselquist2019}, so to push to lower metallicities we must use other means to identify Sgr members.  Fortunately, the Sgr system, including the Sgr stream, possesses a relatively unique orbit that enables Sgr stream members to be readily identified {\it kinematically} from surveys of the MW.  The Sgr stream is also sufficiently close that {\it Gaia} DR2 proper motions \citep{gaiadr2}, APOGEE radial velocities, and spectrophotometric distances can be measured to such a precision that complete 6-D phase space information can be obtained for large samples of candidate stars.  Because a selection of the Sgr stream candidates from the 6-D phase space distribution of APOGEE-observed stars is relatively free from metallicity bias, one can reliably measure chemical gradients along the Sgr stream from the identified stream members.

In this work we perform such a selection of Sgr stream stars based on their 3D positions and velocities relative to the Sgr orbital plane.  We also exploit the fact that APOGEE-2 is now operating in both the Northern and Southern Hemispheres, so that, with the dual hemisphere APOGEE data reported in SDSS DR16 \citep{dr16,jonsson2019}, we can obtain a more complete coverage of both the leading and trailing arms of the the Sgr stream.  With a relatively large sample of Sgr stream members, and the precise multi-element APOGEE abundances, we can also begin probing gradients in chemical abundance ratios along the Sgr stream as well as metallicity gradients.

Section 2 provides an overview of the data and quality restrictions we employ for our study. Section 3 describes the selection criteria applied for identifying Sgr stream stars based on their 3D positions and kinematics within a Galactocentric coordinate system defined by the Sgr orbital plane.  Using the high precision bulk metallicities and chemical abundances that APOGEE measures, in Section 4 we discuss the chemical differences found between the Sgr stream and core in Section 4.1, our assessment of metallicity gradients along the Sgr stream in Section 4.2, the first measurements of non-zero $\alpha$-element abundance gradients along the stream in Section 4.3, and, through the use of an N-body simulation, we collate the data from the two arms to understand the chemical gradients as a function of dynamical age or stripping time in the Sgr stream in Section 4.4.  Section 5 discusses the implications that the measured chemical differences and gradients along the stream have for the chemical structure of the progenitor Sgr galaxy.  Finally, in Section 6 we present our main conclusions.

\section{Data}
\label{sec_data}

The data in this paper come primarily from the APOGEE survey \citep{apogee} and its successor APOGEE-2.  We use the APOGEE data in SDSS-IV DR16 \citep{sdss4,dr16,jonsson2019} that will be made publicly available in December 2019.  This data release includes data taken from both the Northern and Southern Hemispheres using the APOGEE spectrographs \citep{wilson2019} on the SDSS 2.5-m \citep{gunn06} and the 2.5-m du Pont \citep{bv73} telescopes respectively.  The targeting procedure for APOGEE is presented in \citet{zas13,zas17} and \citet{beaton2019}, and details of the data reduction pipeline for APOGEE can be found in \citet{dln15}.  Stellar parameters and chemical abundances are derived from the APOGEE Stellar Parameter and Chemical Abundance Pipeline \citep[ASPCAP;][]{aspcap}, based on the {\sc ferre}\footnote{\url{https://github.com/callendeprieto/ferre}} code, through a similar procedure as in SDSS DR14/15.  For SDSS DR16, ASPCAP has now been updated to use a grid of only MARCS stellar atmospheres \citep{marcs}, rather than Kurucz \citep{kurucz1979,jonsson2019}, and using a new H-band line list from \citet{smith2019} that updates the earlier APOGEE line list presented in \citet{shetrone2015}, all of which are used to generate a grid of synthetic spectra \citep{zamora2015}.

From the full APOGEE sample, we remove stars flagged\footnote{A description of these flags can be found in the online SDSS DR15 bitmask documentation (\url{http://www.sdss.org/dr15/algorithms/bitmasks/})} as having the {\sc starflags}: \textsc{bad\_pixels}, \textsc{very\_bright\_neighbor}, or \textsc{low\_snr} set, or any stars with poorly determined stellar parameters, as may be indicated by the \textsc{aspcapflags}:  \textsc{rotation\_warn} or \textsc{star\_bad}.  { Since we do not expect to detect dwarf stars in APOGEE at the distance of the Sgr stream, we limit our analysis to giant stars by selecting stars with calibrated $\log{g}$ $< 4$}.  In addition we only analyze stars with low velocity uncertainty, $V_{\rm err} \leq 0.2$ \kms, and, when considering chemical abundances in Sections 4 and further sections, we require stars to have S/N $> 70$ { per pixel} spectra to remove stars with lower quality spectra and consequently less reliable ASPCAP-derived stellar parameters and chemical abundances.  We also restrict our chemical analysis in Section 4 and beyond to stars { with effective temperatures warmer than 3700 K}, where APOGEE stellar parameters and chemical abundances are reliably and consistently determined { \citep[for more details on the APOGEE DR16 data quality see][]{jonsson2019}}.  

Since we are interested in kinematically identifying distant Sgr Stream stars, we also remove stars that are associated with known globular clusters based on spatial and radial velocity cuts (except the globular cluster M54 that lies in the Sgr dSph), which helps reduce contamination from globular clusters on similar orbits to Sgr.  While some globular clusters may be associated with Sgr, and therefore participated in its overall evolution \citep{dacosta95,ibata95,dinescu00,bellazzini03,law2010b}, we want to understand the chemical evolution of the main Sgr progenitor, and in any case, globular clusters are contaminated with peculiar chemical pollution differentiating the first generation stars from the second generations that appear to exhibit chemistry unique from the rest of the Galaxy.  We additionally remove the APOGEE fields centered on or near the Large and Small Magellanic Clouds (MCs), which are (unsurprisingly) dominated by the heavy sampling of MC stars and are unlikely to contain Sgr stream stars anyway, because these fields do not lie along the Sgr stream.

We supplement the APOGEE data with proper motions from {\it Gaia} DR2 \citep{gaiadr2} and with spectrophotometric distances calculated with the Bayesian distance calculator \texttt{StarHorse} \citep{santiago2016,starhorse} { using multiple photometric bands, the APOGEE DR16 stellar parameters, and, when possible, parallax priors from {\it Gaia} DR2 \citep{starhorseDR16}.  We use the APOGEE DR16 \texttt{StarHorse} distances \citep{starhorseDR16} rather than those that are calculated more purely from parallaxes, such as the \citet{bailerjones2018} distances, because, as of Gaia DR2, these astrometric distances are primarily driven by priors for sources beyond heliocentric distances of 5 kpc (where the parallax uncertainties become too large), and therefore have large uncertainties ($> 20\%$).  

Such large uncertainties are problematic for identifying Sgr stream stars given that, at the closest point to the sun, the Sgr stream is still beyond 10 kpc away \citep{maj03,kop12}, and motivate using spectrophotometric distances, such as the APOGEE DR16 \texttt{StarHorse} distances, which maintain an internal precision of $\sim 10\%$, even at distances much larger than 10 kpc. While other spectrophotometric distance catalogs are publicly available, we have chosen the APOGEE DR16 \texttt{StarHorse} distance catalog presented in \citet{starhorseDR16} because these distances have been calculated using the new, updated APOGEE DR16 stellar parameters and are available for almost all stars in APOGEE DR16, including the $\sim 170,000$ stars added since the last public data release.  Thus, the \texttt{StarHorse} distance catalog covers our APOGEE sample more completely and self-consistently than other publicly available spectrophotometric distance catalogs that are limited to smaller APOGEE data releases, older versions of the ASPCAP-derived stellar parameters, or stellar parameters derived from other, unassociated data sets \citep[e.g.,][]{wang2016,sanders2018,hogg2019}.}

A small fraction of the stars in our sample have \texttt{StarHorse} distances that are flagged with poor solutions (due to having poor or high infrared extinction, or too few stellar models from which to estimate distances), so we excise these stars from our sample.  { Out of the 437,485 unique APOGEE targets, 256,275 are giants that satisfy our spectroscopic quality restrictions, and from which we remove:  2,581 giants because they are identified as globular cluster members, 8,977 giants that fall in fields around the MCs, and finally 2,595 remaining stars with flagged \texttt{StarHorse} distances.}  After applying these quality cuts, our cross-matched sample of APOGEE observed stars with {\it Gaia} measurements and \texttt{StarHorse} distances amounts to 242,122 giants having measured stellar parameters, chemical abundances, radial velocities, proper motions, and distances, from which we identify Sgr stream candidates.

\section{Tracing the Sgr Stream}

\subsection{Selecting Sgr Stream Candidates}
\label{sec_3.1}

We want to identify Sgr stream members from APOGEE based on their location and kinematics, and now, with the high precision proper motions available from {\it Gaia} DR2 and spectrophotometric distances from \texttt{StarHorse} that are relatively precise even out to large distances, we can find members using full 3D spatial velocities. We calculate the Galactocentric coordinates for our cleaned and cross-matched APOGEE sample, using \texttt{StarHorse} distances assuming $R_{\rm GC, \odot} = 8.122$ kpc \citep{gravity}.  We then include the APOGEE radial velocities and the {\it Gaia} DR2 proper motions to calculate the 3D heliocentric spatial velocities of these stars using the prescription in \citet{js87}, and convert these to Galactocentric space velocities assuming a total solar motion of $(V_{r}, \ V_{\phi} \ V_{z})_{\odot} = (14, 253, 7)$ km s$^{-1}$ in the right-handed velocity notation \citep{schonrich2010,schonrich2012,hayes2018c}.

Because the Sgr stream arches across the sky in a near great circle, it has been historically possible to define relatively precisely the orbital plane of the Sgr system without kinematics \citep{maj03}.  While we can use the Galactocentric positions and velocities of stars within our sample to identify Sgr stream stars by their general net motion, we can make an even more careful selection of these members by considering their motion with respect to the very well-defined Sgr orbital plane.  Therefore, we take the Galactocentric positions and velocities that we have calculated and rotate them into the Sgr orbital plane according to the transformations described in \citet[][here we use the definition of the Galactocentric Sgr coordinates where $\Lambda_{\rm GC} = 0$ is set at the Galactic midplane, sometimes referred to as the $\Lambda_{4}$ coordinate system]{maj03}.\footnote{See also the publicly available code that can be used to perform transformations into the Sgr coordinate systems at \url{http://faculty.virginia.edu/srm4n/Sgr/code.html}.}  This produces a set of position and velocity coordinates (which are most usefully expressed in Cartesian or cylindrical forms) relative to the Sgr orbital plane, rather than to the plane of the Galaxy, but still centered on the Galactic center.

Rather than using a model to predict the location and kinematics expected of Sgr stream stars, we want to use a data-driven selection of these stars, and can then compare them to models as further verification of their membership status.  To first order, we can expect that Sgr stream stars should have conserved their orbital angular momentum, and to the accuracy of our data, the orbital angular momentum of Sgr stream stars within our sample should be the same as the orbital angular momentum of known members of the Sgr dSph.  APOGEE has observed a considerable number of stars in the Sgr core \citep{majewski2013,has17}, which we can use to establish the range of orbital angular momenta of the core, and use that range to select stream candidates.

\begin{figure*}
  \centering
  \includegraphics[scale=0.4,trim = 1.5in 0.75in 1.5in 1.25in, clip]{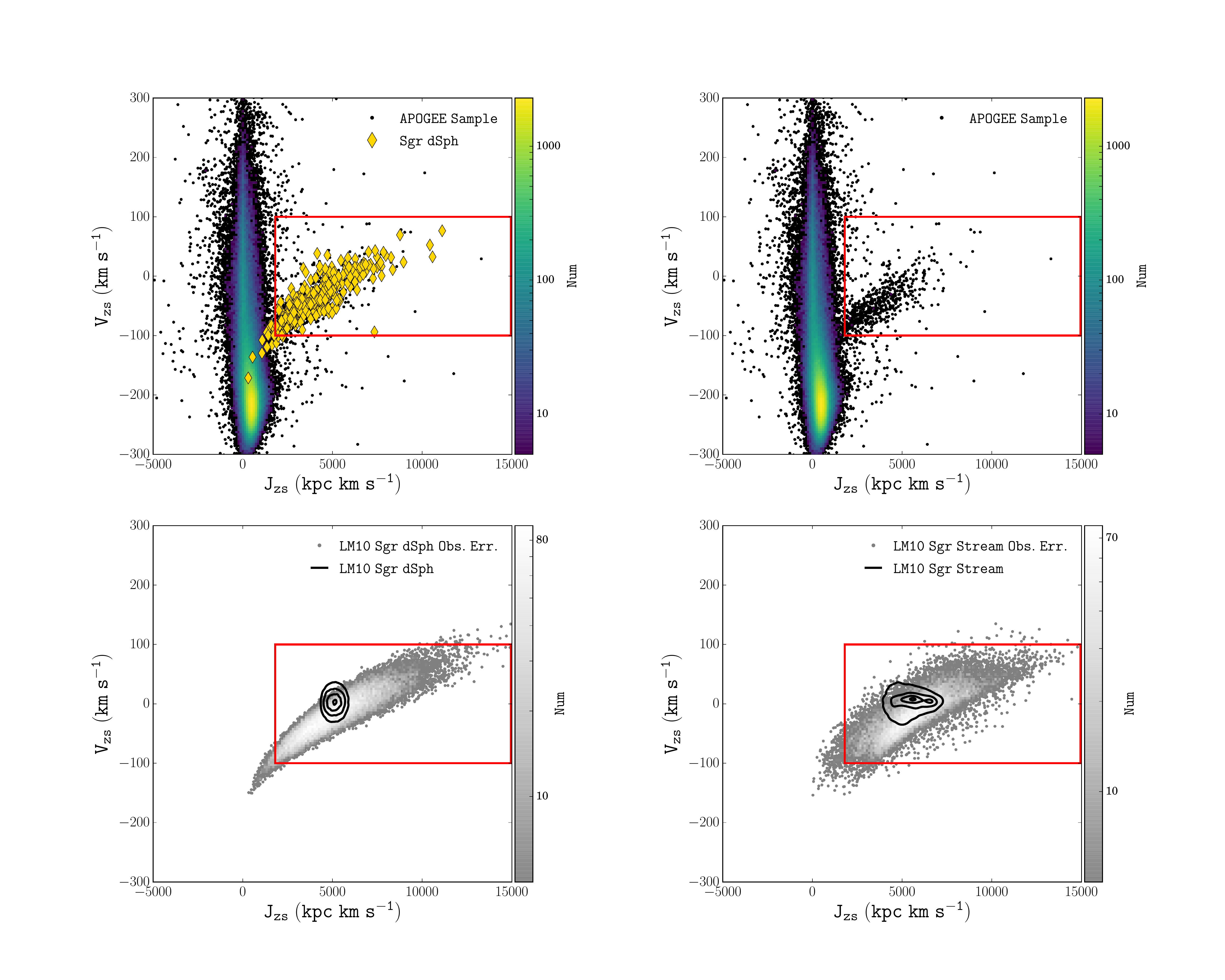}
  \caption{{ Velocity of stars (top panels) or \citetalias{lm10} star particles (bottom panels)} perpendicular to the Sgr orbital plane, $V_{\rm z,s}$ vs. their angular momentum about the axis perpendicular to the Sgr orbital plane, $J_{\rm z,s}$.  (Top panels) APOGEE observed stars with {\it Gaia} DR2 proper motions and \texttt{StarHorse} distances are shown (black points, and 2D histogram for densely populated regions of this space), and the known members of the Sgr core in APOGEE from \citet{majewski2013} and \citet{has17} are highlighted (gold diamonds in the left panel only).  The red box illustrates our initial selection of Sgr stream candidates in this parameter space, and those candidates are shown more clearly in the right panel. { (Bottom left panel) The distribution of \citetalias{lm10} Sgr dSph particles (i.e., particles that are still bound in the model) in this 
  projection of 
  phase space (black contours, containing 95\%, 68\%, 32\%, and 5\% of the particles, from the outside-in), are shown over top of simulated observations of these particles when they are measured with random 10\% distance errors (gray points and 2D grayscale histogram), typical of our \texttt{StarHorse} distance uncertainties.  (Bottom right panel) Same as the bottom left panel, but now illustrating the effect of 10\% distance uncertainties on the distribution of \citetalias{lm10} Sgr stream particles (particles that became unbound within the last three Sgr pericenter passages).}}
  \label{angmom_vzs}
\end{figure*}

Because the Sgr system is relatively well confined to the nominal Sgr orbital plane \citepalias[modulo possible precession of the orbital plane; \citealt{law2005};][]{lm10}, we should expect that stars in the Stream and the dSph to have conserved the same angular momentum (within our uncertainties), and should not have large velocities perpendicular to the orbital plane.  This concept serves as the main selection criteria we employ to select stars in the Sgr dSph and Stream system.  We therefore compute the specific angular momentum of stars in our sample along the z-direction of our Galactocentric Sgr coordinates, $J_{\rm zs} = R_{\rm GC,s} \times V_{\rm \phi,s}$ (i.e., the angular momentum perpendicular to the Sgr orbital plane), and in Figure \ref{angmom_vzs} show the angular momentum of stars in our sample in this direction versus their velocity perpendicular to the Sgr orbital plane ($V_{\rm z,s}$).  Because the Sgr orbital plane is nearly perpendicular to the Galactic plane, most of the APOGEE sample (which is dominated by stars in the disk of the MW near the sun) are rotating with the Galactic disk out of the Sgr orbital plane in the direction of $-V_{\rm z,s}$, and typically have low velocities perpendicular to or radially within the disk of the MW, so they have low velocities in the Sgr orbital plane, and thus a low angular momentum along the direction perpendicular to the Sgr orbital plane.

Known Sgr dSph members from \citet{has17}, however, show a relatively large $J_{\rm z,s}$, as seen in Figure \ref{angmom_vzs}, albeit with a wide spread due to distance uncertainties, but a relatively small velocity perpendicular to the Sgr orbital plane, and identify a range in phase space where we would expect Sgr stream stars to lie.  The correlation between $J_{\rm z,s}$ and $V_{\rm z,s}$ in the Sgr core is an artifact of distance uncertainties inflating/deflating the velocity and angular momenta of core members, because the $V_{\rm z,s}$ and $V_{\rm \phi,s}$ in the direction of the Sgr core predominantly come from proper motions, which are nearly constant across the Sgr core, so a spread in distances will produce a correlated spread in $V_{\rm z,s}$ and $V_{\rm \phi,s}$, and thus between the $V_{\rm z,s}$ and $J_{\rm z,s}$ in the core.  

{ The bottom panels of Figure \ref{angmom_vzs} show particles from the \citetalias{lm10} model in the $J_{\rm z,s} - V_{\rm z,s}$ plane, and compare their distribution in this
projection of 
phase space when measured perfectly (i.e., with no distance errors) versus how they spread when measured with random 10\% distance errors, typical of those in our sample.  These simulated observations demonstrate the affects of distance errors alone, yet appear to mimic the observed correlations seen in our distribution of Sgr dSph members.  The simulations also illustrate that the stream will, as expected, cover a similar region of this parameter space as the Sgr dSph core, and further justifies that the range of $\rm Z_{s}$ angular momenta and velocities of known Sgr dSph members can indicate where we may find Sgr stream candidates.}

To reduce MW contamination, we use a relatively conservative cut in $J_{\rm z,s}$ to select Sgr system candidates, selecting stars with  $J_{\rm z,s} > 1800$ kpc km s$^{-1}$, and remove stars with velocities perpendicular to the Sgr orbital plane $|V_{\rm z,s}| > 100$ km s$^{-1}$ to isolate only those stars with a low velocity perpendicular to the Sgr orbital plane.  To clean out stars that deviate too far from the Sgr orbital plane, we additionally remove any stars that are at Sgr plane latitudes $|B_{\rm GC}| > 20^{\circ}$.  We also remove stars that are within a heliocentric distance of 10 kpc, since the Sgr Stream is known to not come this close to the Sun's position in the MW \citep{maj03,belokurov2014,hernitschek2017}.

\begin{figure*}
  \centering
  \includegraphics[scale=0.375,trim = 1.9in 0.3in 1.9in 0.2in]{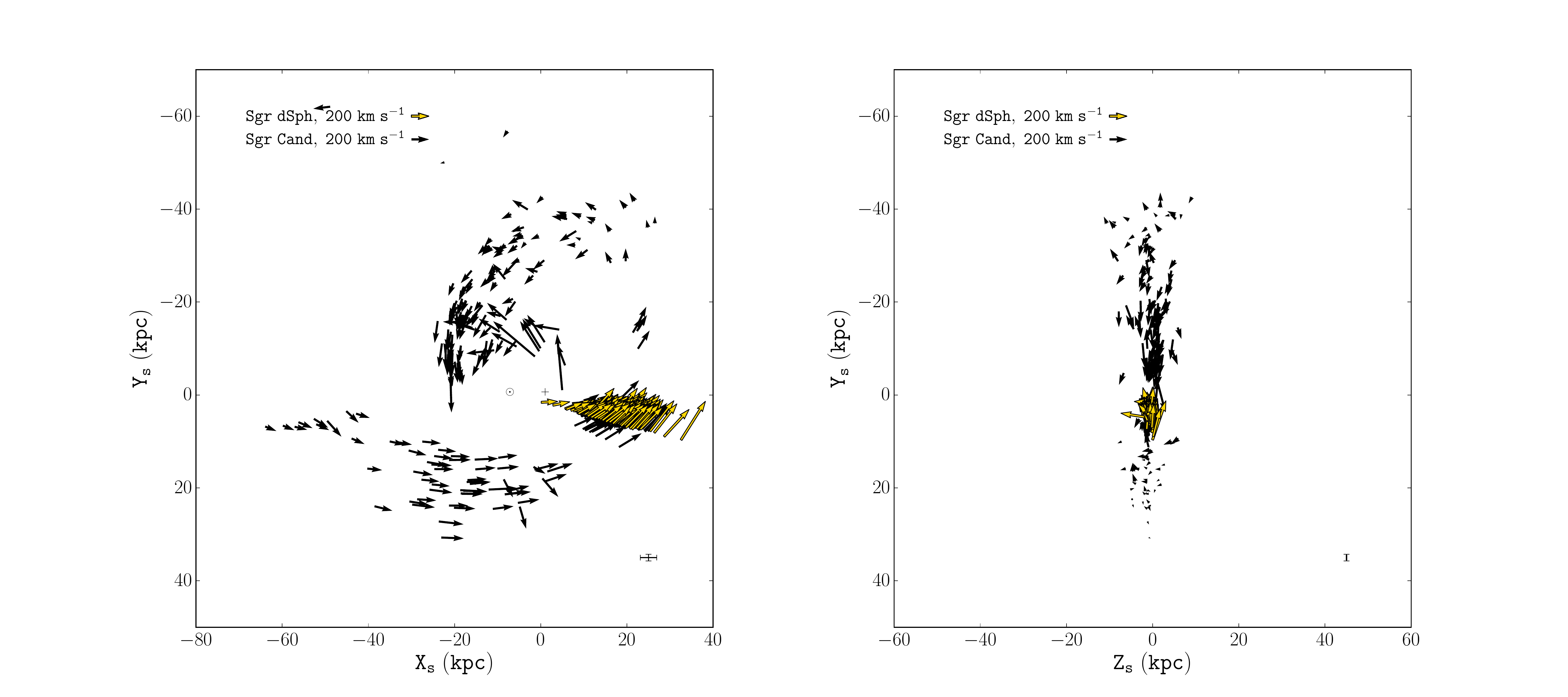}
  \caption{Velocity plot of the Galactocentric distribution of Sgr stream candidates selected as described in Section \ref{sec_3.1} (black arrows), along with known members of the Sgr core \citep[gold arrows; ][]{has17}, as projected onto the Sgr orbital plane of $Y_{\rm s}$ vs. $X_{\rm s}$ and the Galactocentric Sgr $Y_{\rm s}$ vs. $Z_{\rm s}$ plane.  { The median $1\sigma$ uncertainty on these positions is shown bottom left-hand corner (the orientation of these uncertainties is dominated by the Sgr dSph core and the typical orientation of uncertainties for a given star will have maximal uncertainty parallel to the sun-star direction).}  The arrows depict the direction and magnitude of the velocity of these stars in this plane.  For reference, the location of the sun and the  Galactic center are marked (as an $\odot$ and $+$ symbol respectively).  While there appears to be some minor contamination from halo field stars, the bulk of this sample of Sgr stream candidates appear to follow the direction of the Sgr stream with coherent change in the magnitude of velocities along the stream as orbits reach apocenter or pericenter (both in magnitude and direction).}
  \label{cand_xs_ys}
\end{figure*}

\subsection{Removing Halo Contamination}
\label{sec_3.2}

This initial selection of Sgr stream candidates is shown in Figure \ref{cand_xs_ys}. The Sgr stream stands out prominently, as the arc of leading arm stars above the disk, $Y_{\rm s} < 0$, and the curve of trailing arm stars below the disk, $Y_{\rm s} > 0$, but is still contaminated by what appear to be remaining halo stars, seen as stars with peculiar velocity vectors, which we want to remove.  This potential halo contamination comes in two flavors:  (1) stars moving in directions inconsistent with the photometrically implied motion of the Sgr stream (i.e., stars that move nearly perpendicular to the direction of the stream at their location), and (2) stars that are moving in the correct direction, but are still too close to the Sun to be consistent with the location of the stream (despite attempting to remove such contamination by removing stars within 10 kpc of the Sun), even when accounting for distance uncertainties. 

Most of the contamination appears to be above the MW disk ($Y_{\rm s} < 0$), and is particularly noticeable at $Y_{\rm s} \sim -10$ kpc, where there is a spread in the $X_{\rm s}$ distribution of our Sgr stream candidates of about 30 kpc, ranging from $X_{\rm s} \sim -30$ kpc to 0 kpc.  Because the distance to the stream is known to be $\sim 20$ kpc or more in this area of the sky \citep{belokurov2014,hernitschek2017} the spread of stars between  $X_{\rm s} \sim -15$ to $5$ kpc and $Y_{\rm s} \sim -20$ to $-10$ kpc are likely to be halo contamination, because they are too close to the Sun.

While some of these stars have motions that are in the correct direction to be consistent with the Sgr stream, even if their {\sc StarHorse} distances were underestimated, placing them at the distance of the Sgr stream, but keeping their observed proper motions and radial velocities would inflate their space velocities too high for them to be consistent with the rest of the Sgr stream candidates in our sample.  We therefore remove the stars that are too close to the Sun, and are only left with potential halo contamination that is not moving in the correct direction of the stream.

The dominant contributors of stream stars above and below the MW disk midplane are the leading and trailing arms respectively.  Therefore, we would expect stream members to be moving along the direction of the respective arms when we consider stars above and below the disk.  Because we imposed an angular momentum requirement to select our Sgr stream candidates, the stars moving in directions that are inconsistent with the stream around them are primarily stars moving perpendicular to the bulk of our candidate sample. However, the arms of the Sgr stream are thought to cross each other, both above and below the disk, and this crossing could yield a smaller set of candidates from the less dense arm in that Galactic hemisphere that move perpendicular to the stars from the more densely populated arm.  We want to consider whether we are actually identifying any such stars, or if the stars with peculiar motions are instead contamination from the MW halo.  

In the Northern Galactic Hemisphere, above the MW disk ($Y_{\rm s} < 0$), the leading arm is the more densely populated arm of the Sgr stream, and in the left panel of Figure \ref{cand_xs_ys} we do see some stars that are moving perpendicular to the remainder of our stream candidates in this region.  The \citetalias{lm10} model predicts that the trailing arm should cross the leading arm above the disk at $(X_{\rm s}, \, Y_{\rm s}) \sim (-20, \, -20)$ kpc. However, the position of this crossing in the \citetalias{lm10} model is very sensitive to the shape (and possible time-variance) of the MW gravitational potential, and recent studies suggest that the trailing arm instead crosses the leading arm at a point much further above the plane ($\sim 50$ kpc), or may pass over it entirely \citep{hernitschek2017,sesar2017}.  This means that above the MW disk the trailing arm lies in regions where the density of APOGEE targets (and stream candidates) is much lower, and the few stars we see moving perpendicular to the rest of our sample are likely halo contamination.

The crossing of the leading and trailing arms below the MW disk has remained somewhat elusive, with only a few studies suggesting they have observed a few stars in the leading arm below the disk \citep{majewski2004,chou2007,carballobello2017}, but there is still no convincing trace of the extent of the leading arm after plunges through the crowded and dust extinguished MW plane.  We do find four stream candidates that are moving perpendicular to the bulk of our sample below the disk with roughly the correct position and velocity to be in the leading arm, but given the low number of these stars it is hard to confidently associate them with the Sgr stream.  To provide a conservative sample of Sgr stream members, we will not include these candidates in our final sample, but we do note that they may be {\it bona fide} Sgr Stream members belonging to the leading arm.

\begin{figure*}
  \centering
  \includegraphics[scale=0.4,trim = 1.25in 0.in 1.25in 0.in, clip]{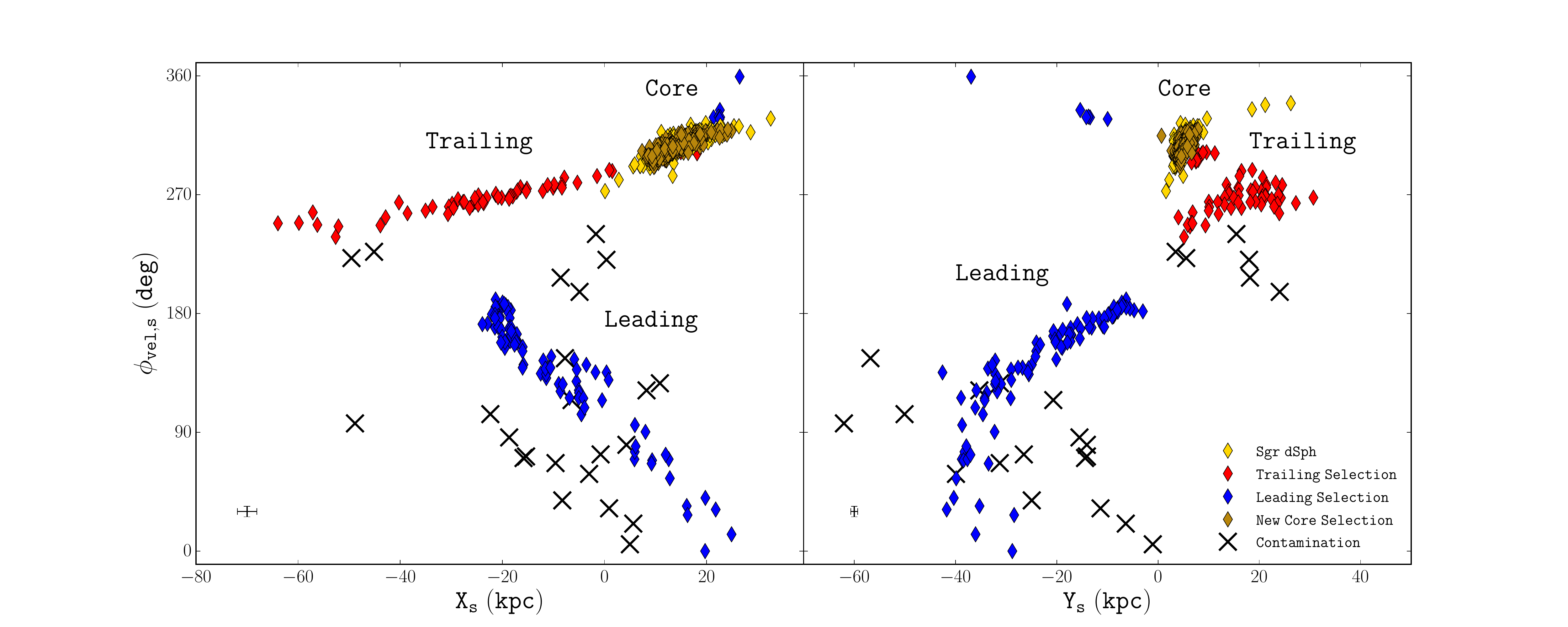}
  \caption{Orbital velocity position angle, $\phi_{\rm vel,s}$, vs. $X_{\rm s}$ (left panel), and vs. $Y_{\rm s}$ (right panel) of the Sgr stream candidates that have been identified as likely contamination (black crosses), those that are likely real members of the trailing arm and leading arm (red and blue diamonds respectively), new core members (dark yellow diamonds), and known Sgr dSph members (as defined in Figure \ref{angmom_vzs}).  { The median $1\sigma$ uncertainties on these positions and angles are shown in the bottom left-hand corner, but note that, as in Figure \ref{cand_xs_ys}, the magnitude of the $X_{\rm s}$ and $Y_{\rm s}$ uncertainties change slightly depending on location and these error bars are most representative of stars located in the Sgr core}.  Stars identified as likely halo contamination move nearly perpendicular to the leading arm at a location where the trailing arm is now established not to cross ($\phi_{\rm vel,s} \sim \pm 90^{\circ}$ from the overdensity of likely Sgr stream members at a given $X_{\rm s}$ or $Y_{\rm s}$ position) and stand away from the Sgr stream locus in one of these two planes, or lie in regions where the leading arm is thought to pass below the MW disk, but the density of Sgr stream stars is low and has not been clearly traced.}
  \label{velangle}
\end{figure*}

To quantitatively remove the aforementioned halo contamination moving in incorrect directions to be members of the Sgr stream, and to remove stars that we cannot confidently associate with the Sgr stream, we assess the candidate stream members' orbital velocity position angle, $\phi_{\rm vel,s} \equiv \arctan(-V_{\rm x,s}/-V_{\rm y,s})$, in the Sgr orbital plane.  This orbital velocity position angle is defined to be zero in the $-Y_{\rm s}$ direction and increasing through the $-X_{\rm s}$ direction, such that as Sgr moves along its orbit it has an increasing orbital velocity position angle.  If the Sgr system were on a perfectly circular orbit, this orbital velocity position angle would be expected to change linearly with Sgr Stream longitude as measured from the Galactic Center, $\Lambda_{\rm GC}$; however, because the Sgr orbit is somewhat eccentric, this relation will vary from linearity.  

In Figure \ref{velangle} we show the orbital velocity position angle, $\phi_{vel,s}$, of the Sgr core members and Sgr stream candidates as a function of their $X_{\rm s}$ and $Y_{\rm s}$ position in the Sgr orbital plane.  Here the leading and trailing arms stand out differently; the leading arm shows a linear distribution in $\phi_{vel,s}-Y_{\rm s}$ at  $Y_{\rm s} \lesssim 0$ kpc that becomes a more tenuous distribution around $Y_{\rm s} \sim 40$ kpc, whereas the trailing arm has a very tight and nearly linear distribution in $\phi_{vel,s}-X_{\rm s}$, but is clumped in the $\phi_{vel,s} - Y_{\rm s}$.  To remove potential halo contamination, we remove any of the Sgr stream candidates that deviate significantly from the stream loci in one of these two planes, and mark which stars have been removed.

\begin{figure}
  \centering
  \includegraphics[scale=0.375,trim = 1.25in 0.3in 1.5in 0.2in]{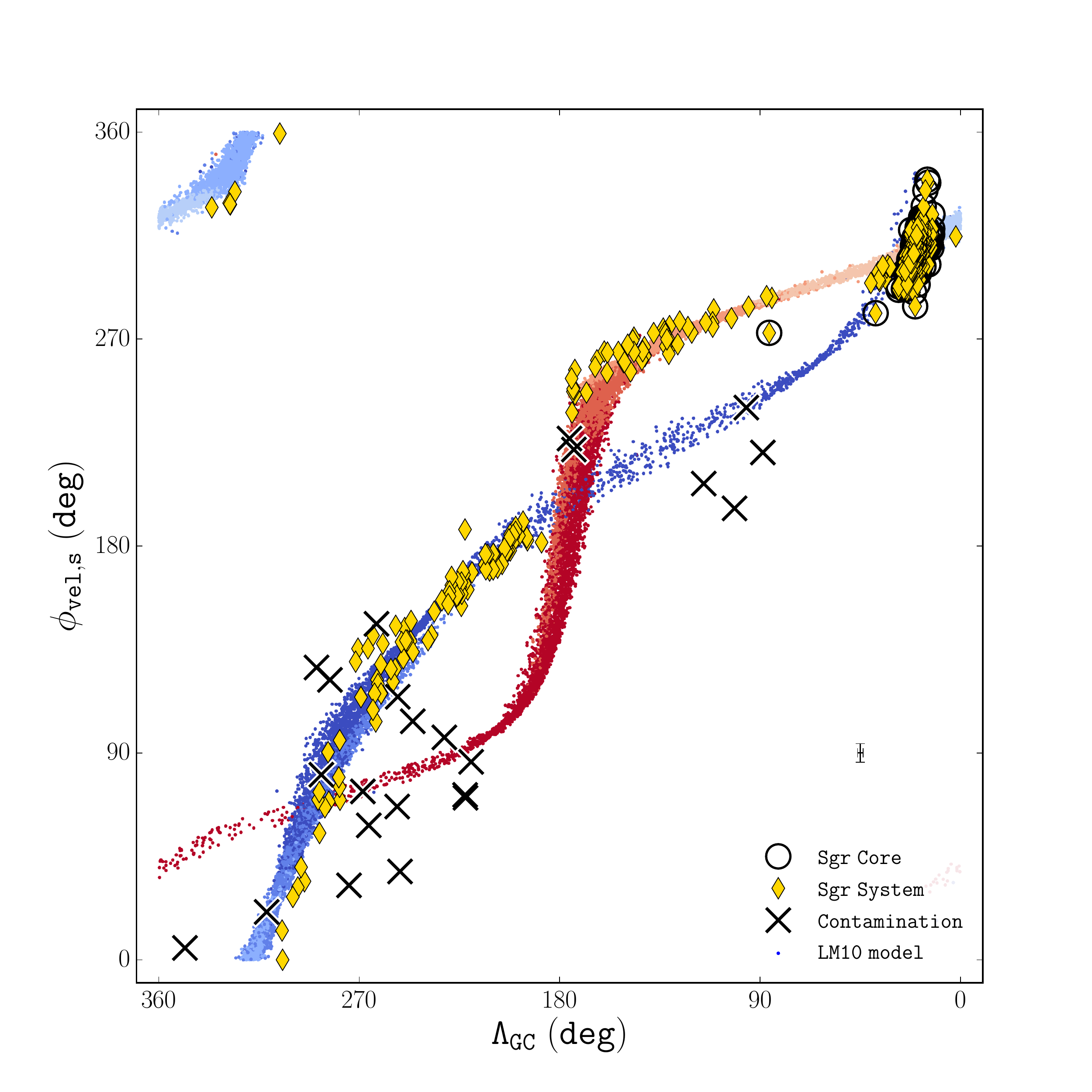}
  \caption{Orbital velocity position angle, $\phi_{\rm vel,s}$, vs. Galactocentric Sgr longitude, $\Lambda_{\rm GC}$ for the final sample of stars selected to be members of the Sgr system (gold diamonds, with the previously known Sgr dSph core members circled in black), and the Sgr stream candidates that were identified as likely halo contamination (black crosses), compared to particles from the \citetalias{lm10} Sgr model (colored points), { with the median uncertainty on these angles shown as the errorbar to the lower right (above the legend)}. The LM10 model points have been colored to identify the leading (blue) and trailing (red) arms with darker saturation corresponding to dynamically older material, stripped off of the Sgr galaxy during earlier pericenter passages. Even though this was not originally a criterion for selection, the Sgr stream members that have been selected via Figure \ref{velangle} closely follow the expectations from the LM10 model, whereas the stars identified as halo contamination deviate more significantly, or lie in regions where the LM10 model is known to not reproduce observations (namely the dynamically oldest parts of the trailing arm).}
  \label{velangle2}
\end{figure}

These Sgr stream member selections on orbital velocity position angle have been applied to remove the stars most inconsistent with our simple hypothesis that the stream should be dynamically coherent, but we can also compare this final selection of stars with the \citetalias{lm10} model to further justify our criteria.  In Figure \ref{velangle2} we compare the $\phi_{vel,s}$ as a function of Sgr longitude as seen from the Galactic Center, $\Lambda_{\rm GC}$, for both our selected sample of Sgr stream stars and the likely MW halo contamination as identified above to predictions from the \citetalias{lm10} model.  

This comparison shows that the final Sgr stream sample is not only very tightly coherent in its distribution, but that it closely follows the predictions of the model, which is reassuring given that this requires a precise combination of the observed distances, proper motions, and radial velocities in the data.  Additionally we see that the stars labeled as likely contamination, by and large, deviate much more significantly from the model.  While there are a few contaminant stars that do line up with the model, they do so along parts of the stream that are poorly modeled/constrained or they physically lie in regions of the halo where the stream (and the rest of our sample) does not pass.

Our final selection identifies 518 new members of the Sgr system in the APOGEE survey, including 133 new Sgr stream stars and 385 new Sgr dSph core stars, and we recover 33 of the 35 metal-rich APOGEE Sgr Stream members found by \citet{hasselquist2019} through chemical tagging.  The advantage here is that our kinematic selection now allows us to push below the [Fe/H] $\sim -1.2$ metallicity that was the limit for the chemical tagging, below which the chemical abundance profile of Sgr begins to blend with that of the MW halo \citep{hasselquist2019}.  The two remaining stream stars that \citet{hasselquist2019} found were not recovered because they lack distances measurements in this APOGEE data release.  

{ To constitute a more complete census of the Sgr system that we analyze throughout the rest of this paper, we combine this sample of new members with (1) the 325 known Sgr dSph stars from \citet{has17} that pass our spectroscopic and distance requirements (299 of which we recover through our Sgr selection; the remaining 26 are excluded by our $J_{\rm z,s} - V_{\rm z,s}$ cuts to avoid MW contamination, as seen in Figure \ref{angmom_vzs})}, and (2) the 33 Sgr stream stars from \citet{hasselquist2019} that we recover. This gives a total sample of { 876} APOGEE observed stars in the Sgr system, the largest sample of Sgr stars with high-resolution spectra to date.  Of the 166 Sgr Stream stars, 103 of them are in the leading arm, and 63 are in the trailing arm.

\begin{figure*}
  \centering
  \includegraphics[scale=0.375,trim = 1.9in 0.3in 1.9in 0.2in]{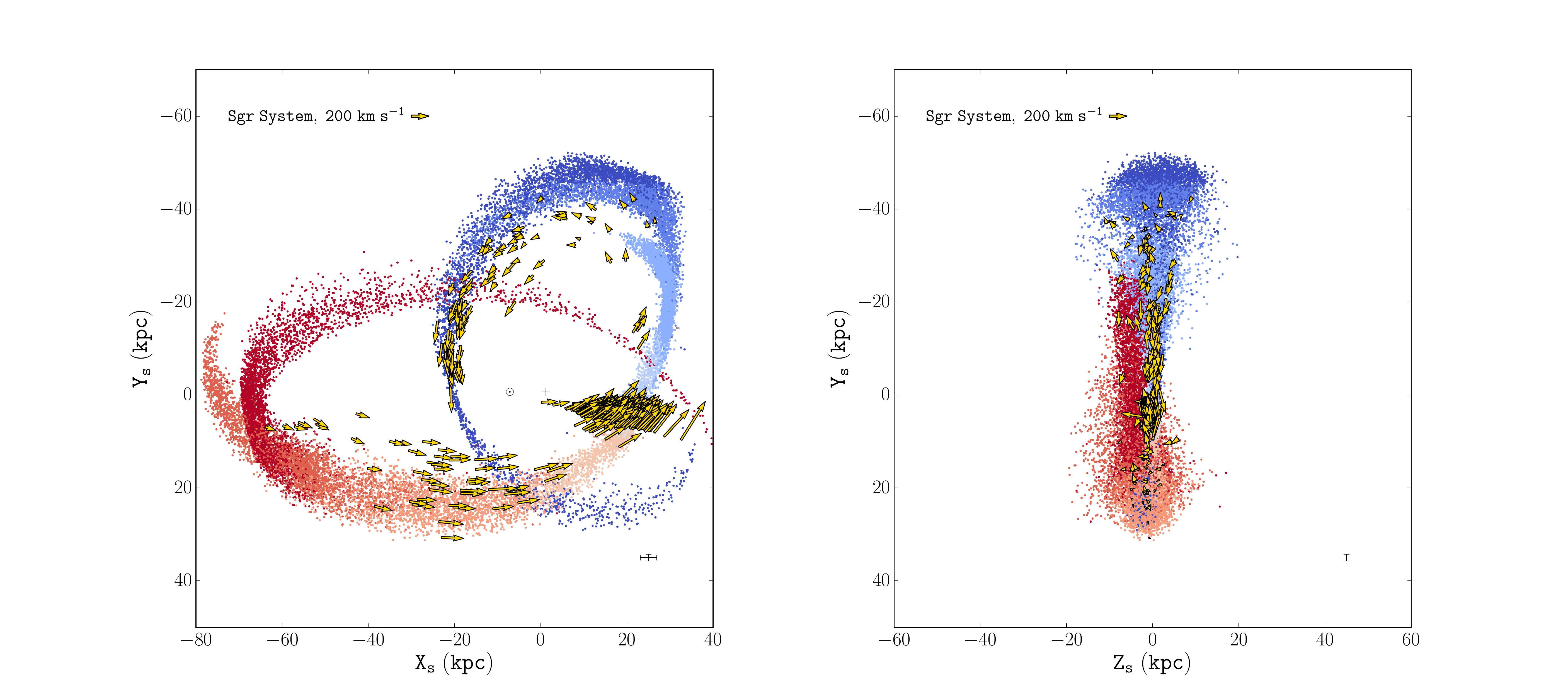}
  \caption{(Left panel) $Y_{\rm s}$ vs. $X_{\rm s}$ and (right panel) $Y_{\rm s}$ vs. $Z_{\rm s}$ projections of the Galactocentric Sgr orbital coordinate system with the distribution of our final sample stars in the Sgr system shown with arrows depicting their projected velocities, and overlaid on particles from the \citetalias{lm10} N-body model, colored as in Figure \ref{velangle2}.  For reference, the location of the sun and the  Galactic center are marked (as an $\odot$ and $+$ symbol respectively), and the typical positional uncertainties are shown as the errorbar in the bottom right-hand corner as in Figure \ref{cand_xs_ys}.  The positions and velocity vectors of the Sgr stream stars in this sample closely follow the distribution from the LM10 model (within the typical $\sim 10\%$ median distance uncertainties ), although the Sgr dSph stars appear to be at systematically closer distances than the model and past distance measurements of Sgr dSph.}
  \label{selection}
\end{figure*}

The distribution of this full Sgr sample throughout the Galactocentric Sgr coordinate system is shown in Figure \ref{selection} overlying the LM10 model of the Sgr Stream pulled off of the main body within the past three pericenter passages ($P_{col} \leq 3$), with arrows illustrating the magnitude and direction of each star's velocity projected onto these planes.  Despite not being selected in accordance with the \citetalias{lm10} model, we can see that on average, our Sgr Stream sample aligns well with the \citetalias{lm10} model in terms of shape and distance for the most part, however, there are two differences:  (1) The width of the Sgr stream in our observed sample appears to be slightly inflated in some places due to distance uncertainties that spread stars along the radial direction from the sun  (although these distances seem to be precise enough to differentiate the narrower width of the leading arm and against wider trailing arm at their points nearest the sun), and (2) there appears to be a difference in the distance scale between the observations and the \citetalias{lm10} model, particularly at the Sgr dSph core and at the apocenter of the leading arm, such that the observed distances are measured closer to the Sun on average.

The median distance to the Sgr core in our sample is about 23 kpc, with a dispersion of $\sigma = 4$ kpc, whereas past studies find slightly larger distances, ranging from 24-28 kpc \citep{monaco2004,siegel2007,mcdonald2013,hernitschek2019}, although we note that our median distance to the Sgr core is still within about 1$\sigma$ of these previously measured distances.  One possible source of our smaller distances to the Sgr core, may be the bulge priors used in calculating \texttt{StarHorse} distances.  To account for the higher density of stars in the Galactic bulge when calculating \texttt{StarHorse} distances, \citet{starhorse} incorporate a prior for stars in the direction of the Galactic Center to lie at distances that place them in the bulge.  Because the Sgr dSph core lies opposite the bulge from the Sun, it lies in a part of the sky where this prior is relevant for MW stars, but it may be skewing the distances of Sgr stars to smaller values.  

However, we can see that the distances to other parts of the stream are also skewed to smaller values than in the \citetalias{lm10} model (by $\sim 15-20\%$).  This may suggest that the \citet{starhorse} values are systematically underestimated at these large distances, or that the \citetalias{lm10} model overestimates the distances to the Sgr system \citep[for which there is some evidence in comparison with the distribution of Sgr stream RR Lyrae, which find slightly closer distances for the apocenter of the leading arm][]{hernitschek2017}.  Regardless, neither possibility should have serious impact on the results that follow, because we use the distance independent heliocentric Sgr longitudinal coordinate system for the remaining analysis.

\begin{figure}
  \centering
  \includegraphics[scale=0.4]{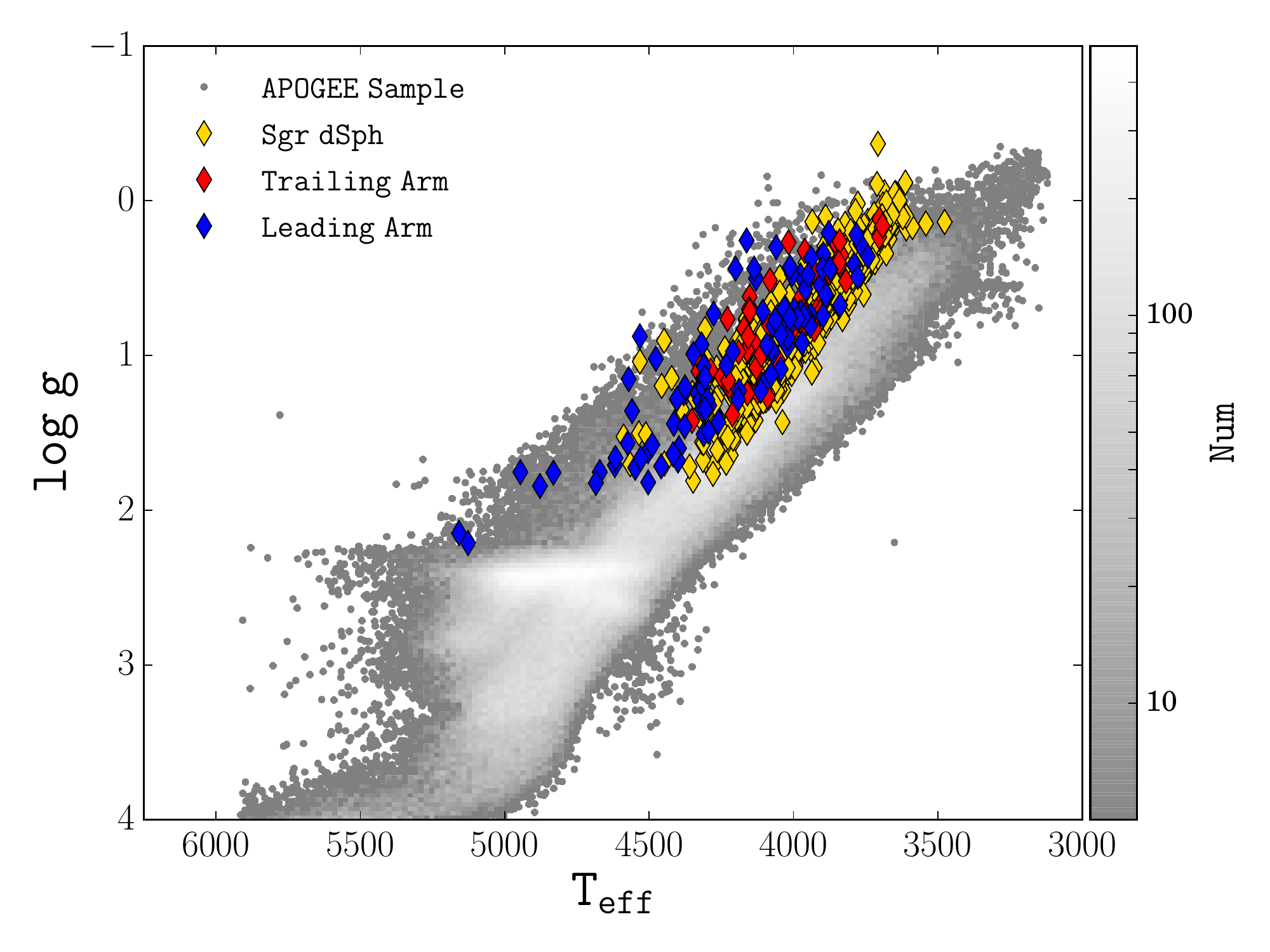}
  \caption{Spectroscopic Hertzsprung-Russel Diagram, showing calibrated log $g$ vs. T$_{\rm eff}$ for our Sgr dSph core (gold diamonds), trailing arm (red diamonds), and leading arm samples (blue diamonds), compared to the remainder of APOGEE giants that meet the quality criteria described in Section \ref{sec_data} (black points and 2D histogram where densely populated).}
  \label{hrd}
\end{figure}

Our final sample of Sgr stars (core and stream) are given in Table \ref{sgr_members}, along with their positions, kinematics, stellar parameters and chemical abundances (for the elements explored in below in Section \ref{sec_chemistry}), as well as, the source of their identification as members of the Sgr system, and their classification as core, trailing arm, or leading arm members.  The stellar parameters for our sample of Sgr stars are shown in the spectroscopic Hertzsprung-Russel diagram in Figure \ref{hrd}, in comparison to the rest of the APOGEE giants that satisfy the quality requirements described in Section \ref{sec_data}.  While we have not applied any temperature cuts prior to this point, as mentioned in Section \ref{sec_data}, for the following analysis in Section \ref{sec_chemistry}, we restrict this sample to calibrated temperatures { warmer than 3700 K}, where APOGEE's stellar parameters and chemical abundances are most reliably and consistently determined for giants.  This only minimally reduces our sample of Sgr stream and core stars, and additionally does not seem to significantly affect our results as discussed in Section \ref{sec_gradients}.

\begin{deluxetable*}{c l p{4.5cm} | c l p{4.5cm}}
\tabletypesize{\scriptsize}
\tablewidth{0pt}
\tablecolumns{3}
\tablecaption{Properties of Sgr Stars \label{sgr_members}}
\tablehead{\colhead{Column} & \colhead{Column Label} & \colhead{Column Description} & \colhead{Column} & \colhead{Column Label} & \colhead{Column Description}}
\startdata
1 & APOGEE & APOGEE Star ID & 24 & R\_cys & Sgr Galactocentric Cylindrical radius, $R_{GC,s}$ (kpc) \\
2 & RAdeg & Right Ascension (decimal degrees) & 25 & V\_xs & Sgr Galactocentric Cartesian $X_{\rm s}$ velocity, $V_{\rm x,s}$ (km s$^{-1}$) \\
3 & DEdeg & Declination (decimal degrees) & 26 & V\_ys & Sgr Galactocentric Cartesian $Y_{\rm s}$ velocity, $V_{\rm y,s}$ (km s$^{-1}$) \\
4 & GLON & Galactic Longitude (decimal degrees) & 27 & V\_zs & Sgr Galactocentric Cartesian $Z_{\rm s}$ velocity, $V_{\rm z,s}$ (km s$^{-1}$) \\
5 & GLAT & Galactic Latitude (decimal degrees) & 28 & V\_rs & Sgr Galactocentric Cylindrical radial velocity, $V_{\rm r,s}$ (km s$^{-1}$) \\
6 & LAMBDA\_sun & Heliocentric Sagittarius Longitude, $\Lambda_{\odot}$ (decimal degrees) & 29 & V\_phis & Sgr Galactocentric Cylindrical rotational velocity, $V_{\rm \phi,s}$ (km s$^{-1}$) \\
7 & BETA\_sun & Heliocentric Sagittarius Latitude, $B_{\odot}$ (decimal degrees) & 30 & SNR & Signal-to-noise ratio of spectrum { per pixel} in APOGEE DR16\tablenotemark{\scriptsize a} \\
8 & LAMBDA\_GC & Galactocentric Sagittarius Longitude, $\Lambda_{\rm GC}$ (decimal degrees) & 31 & Teff & DR16 effective surface temperature (K)\tablenotemark{\scriptsize a} \\
9 & BETA\_GC & Galactocentric Sagittarius Latitude, $B_{\rm GC}$ (decimal degrees) & 32 & e\_Teff & DR16 uncertainty in $T_{\rm eff}$ (K)\tablenotemark{\scriptsize a} \\
10 & Jmag & 2MASS J magnitude & 33 & logg & DR16 surface gravity\tablenotemark{\scriptsize a} \\
11 & Hmag & 2MASS H magnitude & 34 & e\_logg & DR16 uncertainty in $\log g$\tablenotemark{\scriptsize a} \\
12 & Kmag & 2MASS Ks magnitude & 35 & Vturb & DR16 microturbulent velocity (km s$^{-1}$)\tablenotemark{\scriptsize a} \\
13 & Dist & Heliocentric distance (kpc)\tablenotemark{\scriptsize a} & 36 & Vmacro & DR16 macroturbulent velocity (km s$^{-1}$)\tablenotemark{\scriptsize a} \\
14 & e\_Dist & Uncertainty in distance\tablenotemark{\scriptsize a} (kpc) & 37 & [Fe/H] & DR16 log abundance, [Fe/H]\tablenotemark{\scriptsize a} \\
15 & HRV & Heliocentric radial velocity (km s$^{-1}$) & 38 & e\_[Fe/H] & DR16 uncertainty in [Fe/H]\tablenotemark{\scriptsize a} \\
16 & e\_HRV & Radial velocity uncertainty (km s$^{-1}$) & 39 & [Mg/Fe] & DR16 log abundance, [Mg/Fe]\tablenotemark{\scriptsize a} \\
17 & pmRA & Proper motion in RA (mas yr$^{-1}$) & 40 & e\_[Mg/Fe] & DR16 uncertainty on [Mg/Fe]\tablenotemark{\scriptsize a} \\
18 & e\_pmRA & Uncertainty on pmRA (mas yr$^{-1}$) & 41 & [Si/Fe] & DR16 log abundance, [Si/Fe]\tablenotemark{\scriptsize a} \\
19 & pmDE & Proper motion in Dec (mas yr$^{-1}$) & 42 & e\_[Si/Fe] & DR16 uncertainty on [Si/Fe]\tablenotemark{\scriptsize a} \\
20 & e\_pmDE & Uncertainty on pmDE (mas yr$^{-1}$) & 43 & t\_un & Dynamical age estimated from the \citetalias{lm10} model, $t_{\rm unbound}$ (Gyr)\tablenotemark{\scriptsize b} \\
21 & X\_s & Sgr Galactocentric Cartesian $X_{\rm s}$ position (kpc) & 44 & MEMBERSHIP & Identifies stream or core membership\tablenotemark{\scriptsize c} \\
22 & Y\_s & Sgr Galactocentric Cartesian $Y_{\rm s}$ position (kpc) & 45 & STUDY & Identifies the source of membership with the Sgr system\tablenotemark{\scriptsize d} \\
23 & Z\_s & Sgr Galactocentric Cartesian $Z_{\rm s}$ position (kpc) & & & \\
\enddata
\tablecomments{Table 1 is published in its entirety in the machine-readable format.  A portion is shown here for guidance regarding its form and content.}
\tablecomments{Null entries are given values of -9999.}
\tablenotetext{a}{Publicly released in SDSS-IV DR16 \citep{dr16,jonsson2019}}
\tablenotetext{b}{Dynamical ages of -1 refer to stars that are still bound to the Sgr dSph core.}
\tablenotetext{c}{Stars are listed with a membership of ``core,''  ``trailing,'' or ``leading,'' depending on if they are members of the Sgr dSph core, trailing arm, or leading arm, respectively.}
\tablenotetext{d}{Stars are listed with an associated study of ``Has17,''  ``Has19,'' or ``Hay19,'' to denote that they were identified as members of the Sgr system in \citet{has17}, \citet{hasselquist2019}, or this work, respectively}
\end{deluxetable*}

\section{Chemistry Along the Sgr Stream}
\label{sec_chemistry}

\begin{figure*}
  \centering
  \includegraphics[scale=0.55,trim = 1.5in 0.in 1.5in 0.in]{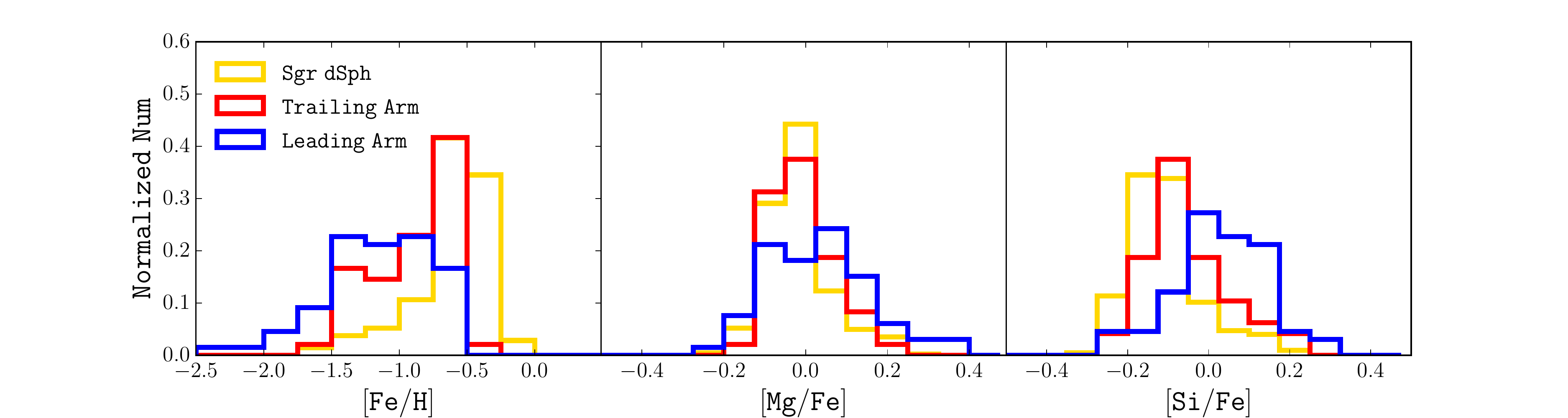}
  \caption{Metallicity, [Fe/H] (left), [Mg/Fe] (middle), and [Si/Fe] (right) distributions of stars in the Sgr dSph core (gold), trailing arm (red), and leading arm (blue), normalized by the number of stars within each sample.  While the metallicity and [$\alpha$/Fe] distributions are generally non-Gaussian, with tails toward lower metallicties and higher [$\alpha$/Fe], the metallicity decreases and the [$\alpha$/Fe] ratio increases when moving from the still bound Sgr dSph stars, through the dynamically younger trailing arm sample, to the dynamically older leading arm sample.}
  \label{mdf}
\end{figure*}

\subsection{Metallicity Differences between Sgr dSph and Sgr Stream}
\label{sec_gradients}

The combination of the identified Sgr stream members and dSph core sample allows us to explore the chemistry of the complete Sgr system, to the extent that the extant stream so far identified represents all stripped populations.  It is immediately evident in Figure \ref{mdf} that the metallicity in each arm of the Sgr stream is lower than that of the Sgr dSph core. The median metallicity of the dSph sample is measured to be [Fe/H]$_{\rm dSph} = -0.57$, whereas the median metallicity of our trailing and leading arm samples are [Fe/H]$_{\rm trailing} = -0.84$ and [Fe/H]$_{\rm leading} = -1.13$ respectively.  Performing a Kolmogorov-Smirnov (KS) test to compare the metallicity distributions of the Sgr core, trailing arm, and leading arm samples indicates a $1\%$ probability that the metallicity distributions of the trailing and leading arm samples are drawn from the same distribution, and a much lower probability ($\ll 1\%$) that either the trailing arm or leading arm samples are drawn from the same metallicity distribution as the Sgr dSph core.

Figure \ref{selection} and Section \ref{sec_tag} illustrate that, by comparison with the \citetalias{lm10} model, our trailing arm sample falls along ranges of the stream predicted to be stripped off of Sgr during the past pericenter passage or two, and is therefore dynamically younger than the leading arm sample that primarily traces material, which was stripped off three pericenter passages ago (i.e., the trailing arm sample traces lighter portions of the model than the deeper saturated parts where the leading arm sample lies).  Comparing the median metallicities of the three samples shown in Figure \ref{mdf} to their relative dynamical ages makes it clear that there is a correlation between dynamical age and metallicity, such that dynamically older material is, on average, more metal poor.  Figure \ref{mdf} also shows the $\alpha$-element distributions of these samples, which is discussed more in Section \ref{alphas_section}.

Assuming that tidal stripping works predominantly outside-in, these dynamical ages roughly trace back to different depths within the Sgr progenitor.  Thus, our leading arm sample would represent the outermost/least bound stars in the progenitor, whereas our trailing arm sample comes from more intermediate radii.  In the presence of an initial metallicity gradient within the Sgr progenitor, we would expect that our leading arm sample would have a lower metallicity population than the stars from our trailing arm, consistent with our findings.

We first note that our [Fe/H] $\sim -0.57$ dex value for the median metallicity in the Sgr dSph is somewhat more metal poor than past measurements around [Fe/H] $\sim -0.4$ \citep[e.g.,][]{monaco2005,chou2007}, and we also find similar differences for the metallicities of the trailing and leading arms.  They are again somewhat more metal-poor than reported by earlier studies of the metallicity in the Sgr stream (particularly in the leading arm), which found the trailing arm to have a metallicity of [Fe/H] $\sim -0.6$ and the leading arm to be in the range of [Fe/H] $ \sim -0.7$ to $-0.8$ in the regions of the stream that we observe \citep{chou2007,monaco2007}.  While the new APOGEE results are closer to the metallicities of the trailing ([Fe/H]$_{\rm trailing}$ $= -0.68$) and leading ([Fe/H]$_{\rm leading}$ $= -0.89$) arms found by \citet{carlin2018}, the latter are still slightly more metal rich than we find.  

While we apply a temperature cut (which would tend to bias our sample to slighly lower metallicities) to our Sgr sample at 3700 K to measure these median metallicities, this has less than 0.01 dex affect on the median metallicity of our core sample (compared to when we include core stars cooler than 3700 K), and only removes one star from our trailing arm stream sample that has a negligible effect on the median metallicity of this sample.  In neither the stream nor the core samples, does this temperature restriction produce a significant enough bias to low metallicities to account for the differences between the values we measure and those reported in past studies.

Instead, the higher metallicities may be because these past studies targeted M giants, which are very effective tracers of the Sgr stream in the MW halo \citep{maj03}, however, M giants are also more metal rich than warmer, bluer K giants (which are more affected by halo contamination and so have received less attention).  Thus, the measurements from these earlier M giant studies were biased to higher metallicities, although the presence of more metal-poor Sgr stream populations was evident through the blue horizontal branch (BHB) and RR Lyrae stars identified in the stream \citep{bellazzini2006,yanny2009,sesar2010}.  

\begin{figure}
  \centering
  \includegraphics[scale=0.45,trim = 0.3in 0.3in 0.3in 0.3in]{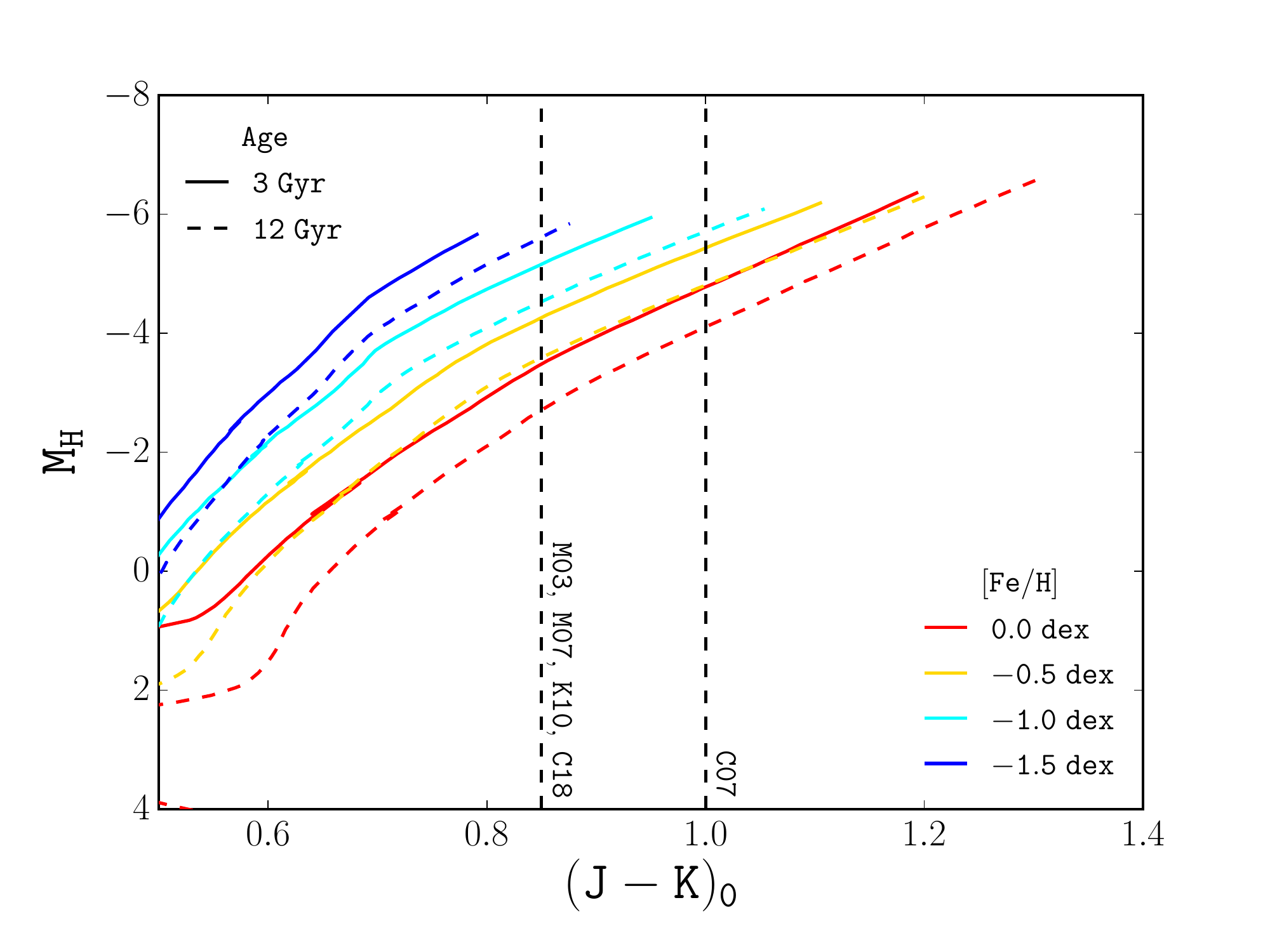}
  \caption{$\rm M_H$ vs. J-K CMD of PARSEC isochrones \citep{parsec} at ages of 3 Gyr (solid lines) and 12 Gyr (dashed lines), which cover the range of ages expected in the Sgr stream, and for metallicities of [Fe/H] $= 0.0$, $-0.5$, $-1.0$, and $-1.5$ dex (red, gold, cyan, and blue respectively).  The black dashed lines show the blue edge of color cuts used in past Sgr stream studies to select M giants, at $(J-K)_0 \geq 0.85$ \citep[][labeled M03, M07, K10, and C18 respectively]{maj03,monaco2007,keller2010,carlin2018}, and at $(J-K)_0 \geq 1.0$ \citep[labeled as C07]{chou2007}.  These color selections bias Sgr stream samples to higher metallicities, because low metallicities, [Fe/H] $\lesssim -1.5$, are almost entirely excluded, and higher metallicity RGBs have a larger stellar parameter coverage with these color limits.}
  \label{kmgiants}
\end{figure}

In Figure \ref{kmgiants} we show the $\rm M_H$ versus $J-K$ CMD for 3 Gyr and 12 Gyr PARSEC isochrones \citep{parsec}, a range of ages that might be expected in the Sgr stream based on population synthesis of the core \citep{siegel2007}, and assuming there are no stars in the stream that were born after Sgr began to be stripped, and a range of metallicities is shown for each age.  The $J-K$ color limits utilized by different studies to select M giants are also shown in Figure \ref{kmgiants}.  Most studies that targeted M giants employed a $(J-K)_0 \geq 0.85$ color cut, either using the Sgr Stream M giant selection from \citet{maj03} or reproducing a similar selection \citep{monaco2007,keller2010,carlin2018}, although \citet{chou2007} targeted even redder stars using a $(J-K)_0 \geq 1.0$ selection.  Comparing these color selections with the PARSEC isochrones, we can see that RGB stars with metallicities [Fe/H] $\lesssim -1.5$ would be excluded almost entirely, regardless of their age, and even at a metallicity of $-1$, only stars at the tip of the RGB would be included in such targeting.  At higher metallicities, a larger span of the RGB is included within the color selection, so, not only do these color selections exclude the lowest metallicity stars, they also bias the sample to higher metallicities, because metal-rich giants can be selected over a larger range of absolute magnitude or stellar parameters.  

In this work, we have a serendipitous targeting of Sgr stream giants throughout the APOGEE survey, rather than a targeting of M giants specifically.  Most of the Sgr stream candidates are in halo fields that have a color selection criterion of $(J-K)_0 \geq 0.3$, although even in designated disk fields the blue edge of target selection is only $(J-K)_0 \geq 0.5$.  Given the much more liberal selection, our sample should be far less affected by metallicity bias than M giant samples, which likely explains our lower metallicities for the Sgr stream.

The Sgr core was targeted more intentionally by APOGEE, and represents a combination of target selections.  The original selection of Sgr core stars in APOGEE-1 is a mix of M giants satisfying the \citet{maj03} selection criteria and known Sgr core members identified by \citet{frinchaboy2012} with a slightly less conservative color selection \citep{zas13}.  These have been supplemented in APOGEE-2 by newly observed Sgr core stars that have a more liberal color selection, $(J-K)_0 \geq 0.5$, and are less biased to high metallicities.  This combination of different selection criteria allows us to probe a similar range of metallicities in the Sgr core to that of our stream sample, however the mix of selection criteria make it difficult to determine what metallicity bias may be present in our Sgr core sample.  We therefore note that there may be some bias in our core sample, although it should be less extreme than a strict M giant color selection, such as that utilized in \citet{maj03}.

As a demonstration of this bias, if we impose a $(J-K)_0 \geq 0.85$ selection to our Sgr stream sample, we increase the median metallicity of our core, trailing, and leading arm samples by around 0.1 dex, to [Fe/H] $= -0.52$, $-0.73$, and $-0.97$ respectively, values that are in better agreement with these M giant studies, especially the recent study from \citet{carlin2018}.  Any remaining differences in the Sgr stream are likely due to stochastic variations or possibly the result of spatial biases that favor particular parts of the stream from which they have been drawn.

\subsection{Metallicity Gradients Along the Sgr Stream}

\begin{deluxetable*}{lrccccc}
\tablewidth{0pt}
\tablecolumns{7}
\tablecaption{Chemical Gradients Along the Sgr Stream \label{grad_table}}
\tablehead{\colhead{} & \colhead{Grad. Method} & \multicolumn{2}{c}{Anchored\tablenotemark{\scriptsize a}} & \multicolumn{2}{c}{Internal\tablenotemark{\scriptsize b}} & \colhead{Dynamical Age\tablenotemark{\scriptsize c}} \\ \colhead{} & \colhead{Units} & \colhead{dex deg$^{-1}$} & \colhead{dex deg$^{-1}$} & \colhead{dex deg$^{-1}$} & \colhead{dex deg$^{-1}$} & \colhead{dex Gyr$^{-1}$} \\ \cline{2-7} \colhead{Element} & \colhead{} & \colhead{Trailing Arm} & \colhead{Leading Arm} & \colhead{Trailing Arm} & \colhead{Leading Arm} & \colhead{Full Stream}}
\startdata
$\rm [Fe/H]$ & & $(2.6 \pm 0.4) \times 10^{-3}$ & $(4.0 \pm 0.3) \times 10^{-3}$ & $(1.2 \pm 0.9) \times 10^{-3}$ & $(1.4 \pm 1.4) \times 10^{-3}$ & $0.12 \pm 0.03$\\
$\rm [Mg/Fe]$ & & $(0.2 \pm 0.1) \times 10^{-3}$ & $(0.5 \pm 0.1) \times 10^{-3}$ & $(0.3 \pm 0.2) \times 10^{-3}$ & $(0.8 \pm 0.5) \times 10^{-3}$ & $0.02 \pm 0.01$\\
$\rm [Si/Fe]$ & & $(0.5 \pm 0.1) \times 10^{-3}$ & $(1.1 \pm 0.1) \times 10^{-3}$ & $(0.2 \pm 0.3) \times 10^{-3}$ & $(0.9 \pm 0.4) \times 10^{-3}$ & $0.04 \pm 0.01$\\
\enddata
\tablenotetext{a}{The gradient measured along each arm of the stream when anchoring the gradient by the chemistry of the Sgr dSph core.}
\tablenotetext{b}{The gradient measured internally within each arm of the stream excluding the chemistry of the Sgr dSph core.}
\tablenotetext{c}{The gradient measured within the Sgr stream as a function of the estimated dynamical ages (i.e., stripping times) of Sgr \\ stream stars and combining both arm samples. See Section \ref{sec_tag} for details.}
\end{deluxetable*}

\subsubsection{Metallicity Gradients in the APOGEE Sample}

\begin{figure*}
  \centering
  \includegraphics[scale=0.7,trim = 2.in 0.in 2.in 0.in]{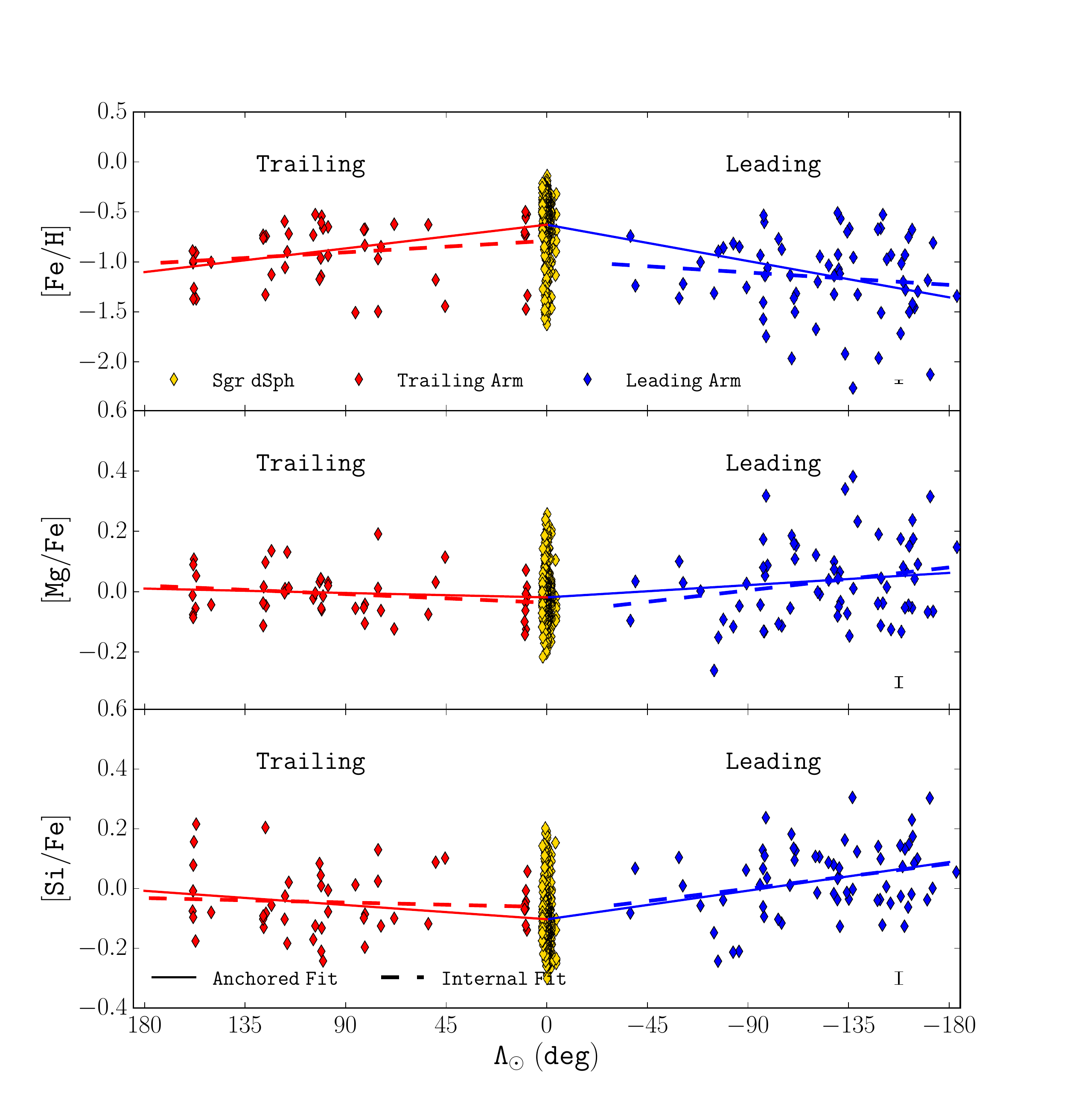}
  \caption{Metallicity, [Fe/H] (top), [Mg/Fe] (middle), and [Si/Fe] (bottom) vs. solar-centered Sgr longitude, $\Lambda_{\odot}$, of Sgr dSph (gold), trailing arm (red), and leading arm (blue) stars.  These distributions have been fit assuming a linear metallicity and abundance gradient with $\Lambda_{\odot}$ through the trailing (red line) and leading (blue line) arms, when anchored by the chemistry of the Sgr dSph core (solid line) and when only measured internally along the stream (dashed line).  The median uncertainty on each elemental abundance in our sample is shown as the errorbar in the bottom right-hand corner to illustrate the typical uncertainties.   The gradients anchored by the chemistry of the Sgr dSph core are flatter along the dynamically younger trailing arm, qualitatively consistent with expectations of tidally stripping a Sgr progenitor galaxy having initial radial metallicity and $\alpha$-element abundance gradients.  The internal gradients within each of the arms are both flatter than the gradients that are measured when anchoring them to the chemistry of the Sgr dSph core, consistent with the limited dynamical age range within the samples of each arm.}
  \label{gradients}
\end{figure*}

The metallicity differences between the Sgr dSph and the trailing and leading arms suggest that there may be metallicity {\it gradients} along the Sgr stream.  In the top panel of Figure \ref{gradients}, we show the metallicity of stars in our sample as a function of their Sgr longitude ($\Lambda_{\odot}$).  Despite the relatively large (astrophysical) scatter in the samples of each arm, there do appear to be gradients in the metallicity along each arm, depending on how the gradients are measured. Considering each separately, we fit the metallicity of the Sgr member stars with a linear trend as a function of Sgr longitude anchored at the Sgr dSph core (by including Sgr stream and core members in the fits) to measure metallicity gradients and their uncertainties { (i.e., the formal error from linear regression, not including the uncertainties on each abundance, since these smaller than the intrinsic dispersion, and therefore do not accurately capture said dispersion)}, which are reported in the first row of Table \ref{grad_table} under the ``Anchored'' gradient measurement method, for the trailing and leading arms respectively.

If the original Sgr progenitor galaxy had a spatial metallicity gradient (with a larger fraction of higher metallicity stars more interior/tightly bound), then on average, with stripping assumed to proceed from the outermost part of the progenitor galaxy to smaller radii, lower metallicity stars would tend to be stripped off of the Sgr progenitor at earlier times and higher metallicity stars would be stripped during successive pericenter passages, and this process would naturally produce gradients along the stream.  Because there is a better energy sorting along the trailing arm, such that debris of different dynamical ages are more spread out along the extent of the trailing arm \citep{chou2007,keller2010,niederste2012}, the stars pulled off during successive pericenter passages are more clearly delineated in the trailing arm than the leading arm.  Therefore, the distribution of dynamical age along the trailing arm is more extended on the sky, so that observationally the metallicity gradient along the trailing arm should be shallower than along the leading arm, as seen here. 

The Sgr core is known to have undergone star formation and enrichment after the tidal stripping of the Sgr progenitor began \citep{siegel2007}, therefore, internal gradients within the tidal debris of the Sgr stream may be more indicative of radial metallicity gradients within the Sgr progenitor than measuring gradients anchored by the present-day Sgr dSph core.  Fitting a linear trend to the metallicity of our trailing and leading arm samples again as a function of Sgr longitude, but now excluding the Sgr dSph core, we find that these internal gradients are much flatter and typically more uncertain than what we find when requiring the stream gradients to be anchored by the Sgr dSph core (see Table \ref{grad_table} under ``Internal'' gradient measurement method column).  In the leading arm, in particular, our sample is consistent (to within 1$\sigma$) with no internal metallicity gradient.

As mentioned above, our leading arm sample (as well as most literature samples of the leading arm) is dominated by dynamically older material that was likely stripped off approximately three pericenter passages ago.  Because the majority of our leading arm sample was stripped around the same time, these stars should be a fairly homogeneous population likely tracing the outermost regions of the Sgr progenitor, even though they are spread out over a large region of the sky.  While this leading arm sample is more metal poor than our trailing arm sample, it covers less breadth in its dynamical age, and therefore has a shallower (or negligible) internal metallicity gradient.  By reference to the \citetalias{lm10} model in Figure \ref{selection}, our trailing arm sample covers a couple pericenter passages in dynamical age, and may therefore have a non-negligible internal metallicity gradient.

\subsubsection{Comparison with the Literature}

Studies measuring metallicity gradients along the Sgr stream in the Literature report a mix of what we have called ``anchored'' and ``internal'' gradients along the Sgr stream.  Considering first the studies that report metallicity gradients anchored by the metallicity of the Sgr core, our metallicity gradients are in good agreement with the results from \citet{keller2010} and \citet{hyde2015}, who find gradients of $(2.4 \, \pm \, 0.3) \, \times \,  10^{-3}$ dex deg$^{-1}$ \citep{keller2010} and $2.7 \, \times \, 10^{-3}$ dex deg$^{-1}$ \citep{hyde2015}\footnote{This is an estimated gradient based on their measurement of a mean core metallicity of [Fe/H] $= -0.59$, which drops to an average metallicity of [Fe/H] $= -0.97$ in a trailing arm sample 142$^\circ$ away from the core}, respectively.  While  \citet{keller2010} exclusively studied the trailing arm, \citet{hyde2015} also measure metallicities for a sample of leading arm stars, however, due to the large metallicity dispersion they found and the metallicity differences between their sample and the leading arm sample from \citet{chou2007}, they do not report any metallicity gradient along the leading arm.

\citet{carlin2012} and \citet{shi2012} both report internal metallicity gradients that they measure, excluding the core, using samples that cover a similar angular extent to that which is covered by our sample.  From low-resolution spectroscopy of stars in six fields along the Sgr trailing arm ranging from $\Lambda_{\odot} = 70^{\circ} - 130^{\circ}$, \citet{carlin2012} find metallicities slightly more metal poor than we find in this range of the stream, but that exhibit a metallicity gradient of $1.4 \times 10^{-3}$ dex deg$^{-1}$ like we find in our sample.  Again, like we find with our trailing arm sample, they suggest that this internal gradient is relatively uncertain, and their sample also seems to be relatively consistent with no internal gradient. 

Using a sample of stars selected from the Sgr stream observed with SDSS DR7 spectroscopy, \citet{shi2012} measured a metallicity gradient $(1.8 \pm 0.3) \times 10^{-3}$ dex deg$^{-1}$ along the trailing arm, similar to ours and \citet{carlin2012} along the trailing arm.  Within uncertainties, the gradient \citet{shi2012} measure along the trailing arm is the same or steeper than the metallicity gradient of $(1.5 \pm 0.4) \times 10^{-3}$ they found along their leading arm sample.  Given the large uncertainty that we have on the internal gradient along the leading arm, this too is consistent with our measurement.

\subsection{$\alpha$-element Abundance Gradients along the Sgr Stream}
\label{alphas_section}

In addition to exploring metallicity gradients, \citet{keller2010} also measured [O/Fe] and [Ti/Fe] abundances for their sample of stream stars, but report little-to-no measureable $\alpha$-element abundance gradients along the trailing arm. This may be a result of their small sample sizes and relatively large abundance ratio dispersion along the trailing arm (which was larger than their measurement uncertainties).

One advantage of the APOGEE database is that it enables the measurement of gradients along the stream homogeneously, accurately (with R $\sim$ 22,500 spectroscopy), and over a relatively continuous and even sampling of both the leading and trailing arms of the Sgr stream, primarily due to the serendipitous targeting of Sgr stream stars throughout the dual-hemisphere APOGEE survey.  Thus, statistically significant abundance gradients along the Sgr stream can be measured for the first time.

In addition to measured metallicity differences between the Sgr dSph core and stream, we find differences in the $\alpha$-element abundance ratios between the Sgr dSph and the streams, and between the trailing and leading arms.  This is illustrated with the examples of Mg and Si \citep[the $\alpha$-elements measured most precisely by APOGEE, which are also reliably measured across the full parameter range of Sgr stars studied here;][]{jon18} and the [Mg/Fe] and [Si/Fe] ratio distributions in the right panel of Figure \ref{mdf}.  The median [Mg/Fe] ratios of the dSph core, trailing arm, and leading arm are [Mg/Fe]$_{\rm dSph} = -0.03$, [Mg/Fe]$_{\rm trailing} = -0.01$, and [Mg/Fe]$_{\rm leading} = +0.03$, respectively and the median [Si/Fe] ratios of the dSph core, trailing arm, and leading arm are [Si/Fe]$_{\rm dSph} = -0.12$, [Si/Fe]$_{\rm trailing} = -0.07$, and [Si/Fe]$_{\rm leading} = +0.03$, respectively.  

Additionally we perform a KS tests on each pair of samples, and for Mg we find that there is a $36\%$ probability that the [Mg/Fe] distribution of the trailing arm is drawn from the same distribution as the Sgr core, a $4\%$ probability that the [Mg/Fe] distributions of the trailing and leading arm are dran from the same population, and finally a much lower probability ($\ll 1\%$) that the [Mg/Fe] distributions core and leading arm are drawn from the same parent population.  In the case of Si, KS tests find a very low, $\ll 1\%$, probability that any of these three samples are drawn from each other's [Si/Fe] distributions, suggesting that these samples are more differentiated in their [Si/Fe] abundance ratios than in [Mg/Fe].

While the $\sim$ 0.06 dex difference in [Mg/Fe] is smaller than the $\sim$ 0.15 dex difference in [Si/Fe] between the Sgr dSph core and leading arm, these differences may suggest a gradient in the detailed chemical abundance patters along the Sgr stream.  The [Mg/Fe] and [Si/Fe] of core and stream stars are shown in the middle and bottom panels of Figure \ref{gradients} respectively, and we measure a gradient from the Sgr dSph core through each arm of the stream to find statistically significant gradients in Mg and Si abundances when anchoring the gradient to the Sgr dSph core, the magnitudes of which are reported along with their uncertainties in Table \ref{grad_table}.

These gradients in $\alpha$-element abundances appear to be associated with the anti-correlation between $\alpha$-element abundance and metallicity along the $\alpha$-shin of the $\alpha$-Fe abundance pattern in the Sgr system \citep{has17,mucciarelli2017,carlin2018,hayes2019inprep}.  This anti-correlation is typical of the chemical abundance patterns of dwarf galaxies and chemical evolution models of such systems, and arises due to a change in the relative contribution of core collapse and Type Ia supernovae (SNe). Therefore, this observed $\alpha$-element abundance gradient along the stream primarily reflects the overall metallicity gradient, but also demonstrates that the material in the streams is also less chemically evolved than the majority of the present-day Sgr dSph core.

As with the internal metallicity gradients within either arm of the stream, we typically find that the internal [Mg/Fe] and [Si/Fe] gradients in Table \ref{grad_table} are flatter or more uncertain, particularly within the trailing arm.  However, unlike the internal metallicity gradients, we find that there is still evidence for an internal $\alpha$-element abundance gradient along the leading arm in both Mg and Si. { This suggests that, despite the shallow internal metallicity gradient along the leading arm, there appears to be some age or population gradient, perhaps even in a single Sgr stripping episode.}

\subsection{Gradients with Dynamical Age}
\label{sec_tag}

While we can study the trailing and leading arms of the Sgr stream separately, in order to build a more complete picture of the Sgr stream  we would ideally want to understand how the metallicity and chemistry of the Sgr stream changes with dynamical age (i.e., stripping).  One way that we can combine the information learned from each of the arms to begin to study the full Sgr progenitor galaxy is by using the \citetalias{lm10} model in concert with our observations. 

The \citetalias{lm10} model records when particles were stripped off of the Sgr galaxy and tracks this information to the present-day location of those particles.  We can, therefore use the model particles at their present-day location in the sky to obtain a rough understanding of the dynamical age of observed Sgr stream stars in the same area of the sky, although we do note that this will, therefore, mean that the below results are model dependent on the \citetalias{lm10} model, and may vary if another model were used.  

To do so, for each observed Sgr star, we find all of the model particles within 5$^{\circ}$ of that star on the sky that have been stripped off within the past three pericenter passages \citepalias[the portion of the model that is best matched to observations;][]{lm10}, i.e., P$_{\rm col} \leq 3$ in the \citetalias{lm10} notation, and paint the median dynamical age of these model particles onto the observed Sgr stream stars.  To match stars with nearby model particles, we only include those particles that cover parts of the stream for which we have observed stars.  This is particularly important for regions of the sky where the arms of the Sgr stream overlap.  For example, we do not include particles from the trailing arm that lie at $Y_{\rm s} < 0$ kpc, since we do not have any stars in this part of the stream.  Because they lie in the same part of the sky as the more densely populated leading arm, the former particles could erroneously push up the median dynamical age of particles in that region of the sky and bias our ages.

\begin{figure*}
  \centering
  \includegraphics[scale=0.7,trim = 2.in 0.in 2.in 0.in]{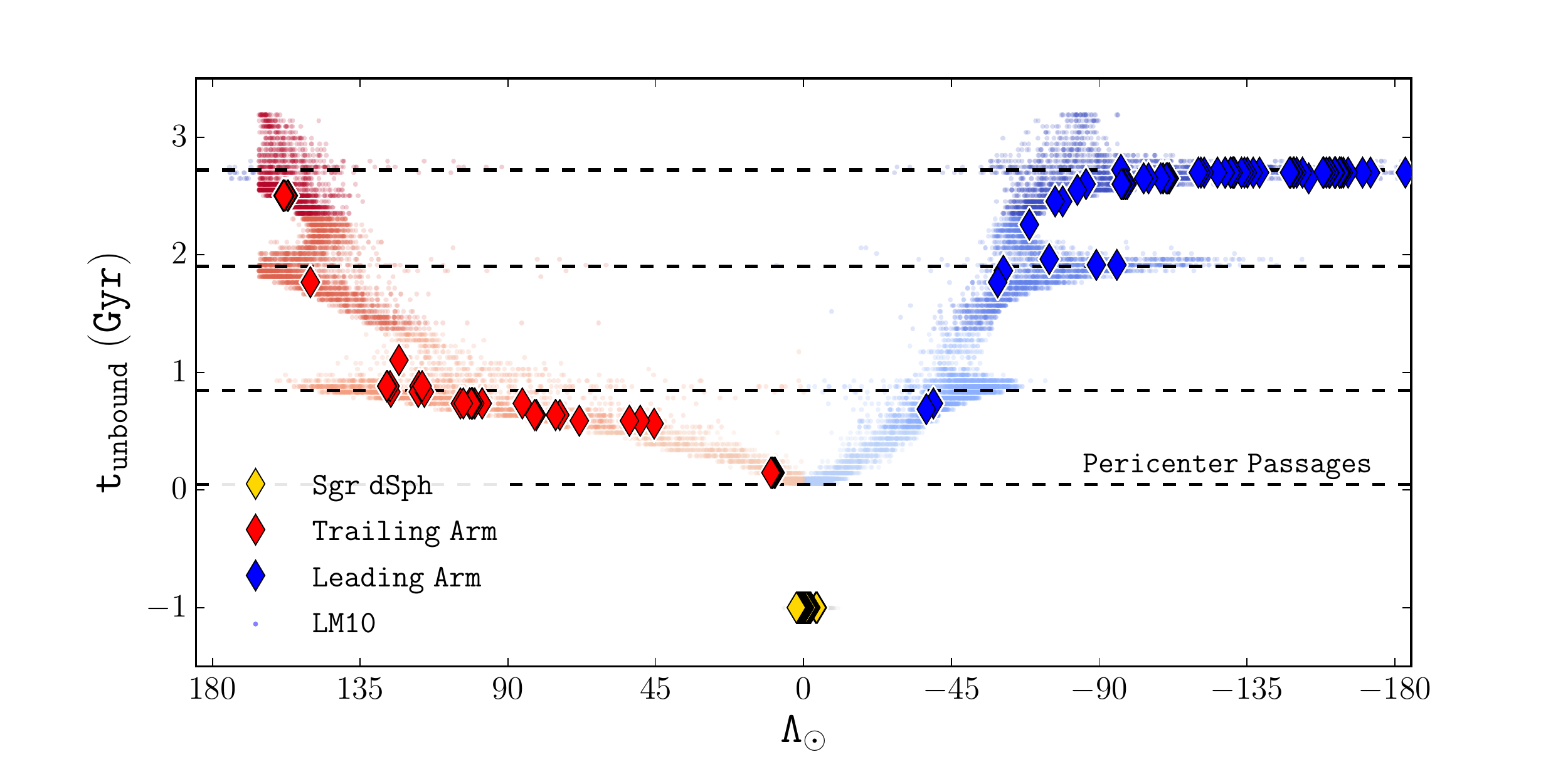}
  \caption{Dynamical age, $t_{\rm unbound}$, vs. solar Sgr longitude, $\Lambda_{\odot}$, for our Sgr sample colored as in Figure \ref{gradients}, and adopting the median dynamical age of Sgr stream particles from the \citetalias{lm10} model (small colored points) within 5$^{\circ}$ of each star (stars still bound to the Sgr dSph are assigned a dynamical age of $-1$ Gyr).  The model particles are colored as in Figure \ref{velangle2}, with a slight transparency to illustrate the more densely populated regions.  Pericenter passages in the model are marked (black dashed lines) and highlight how much of the material in the Sgr stream is pulled off during these episodes and became spread over a large part of the sky.}
  \label{tub_lambda}
\end{figure*}

\begin{figure}
  \centering
  \includegraphics[scale=0.47,trim = 0.in 0.2in 0.in 0.2in]{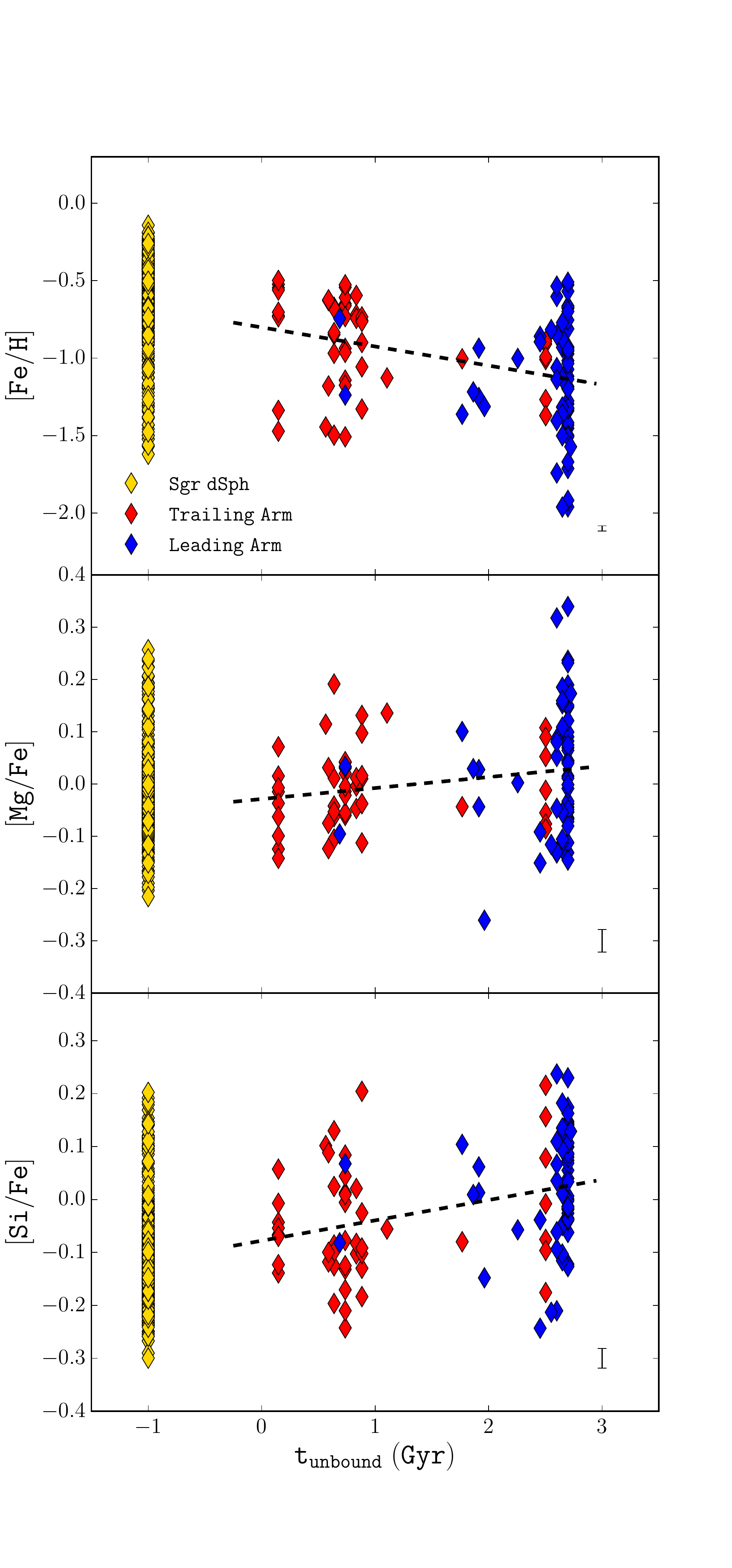}
  \caption{Metallicity, [Fe/H] (top), [Mg/Fe] (middle), and [Si/Fe] (bottom) of stars in the Sgr system observed by APOGEE vs. $t_{\rm unbound}$, the estimated dynamical age each star as described in the text.  The median uncertainty on each elemental abundance in our sample is shown as the errorbar in the bottom right-hand corner to illustrate the typical uncertainties.  Stars from the Sgr dSph, trailing arm, and leading arm are colored as in Figure \ref{gradients}. Collapsing information from both arms into one dimension, we can see a coherent gradient in metallicity, [Mg/Fe], and [Si/Fe] with the expected dynamical age of stars in the Sgr stream, fit by a linear function (black dashed line).}
  \label{tub_tag}
\end{figure}

The estimated dynamical ages for our Sgr sample are shown in Figure \ref{tub_lambda} as a function of their solar Sgr longitude, superposed over the \citetalias{lm10} model from which these ages were drawn. This Figure highlights that our trailing arm sample is dynamically younger than our leading arm sample.  Figure \ref{tub_lambda} also illustrates that, despite our trailing and leading arm samples each covering a large range of Sgr stream longitudes, the sample of stars within each arm were predominantly stripped off from the Sgr galaxy around the same time, and typically around the time of a Sgr pericenter passage.  By considering either arm separately, we are not probing large ranges of the dynamical history of the Sgr stream, and, therefore any either particular arm biases our view of the chemical evolution history of Sgr itself.

Painting dynamical ages onto our observed sample of Sgr stream stars allows us to fold the information from each of the arms of the Sgr stream into one dimension and consider both arms together.  While our total Sgr stream sample does still cluster around the dynamical ages of $\sim 800$ Myr and $\sim 2.7$ Gyr (corresponding to the pericenter passages on Sgr's last orbit and three times ago), we still build a more complete picture of the chemical history of Sgr than we would by focusing on the core or either particular arm alone.  The metallicity, [Mg/Fe], and [Si/Fe] ratios of observed Sgr stars are shown as a function of their estimated dynamical age in Figure \ref{tub_tag}, and reveal coherent gradients of decreasing metallicity and increasing $\alpha$-element abundance with increasing dynamical age as given in the final column of Table \ref{grad_table}.

These gradients with dynamical age are still only a coarse proxy for gradients within the original Sgr galaxy, but do provide evidence that such gradients existed.  The fact that there is also a gradient in $\alpha$-element abundances using [Mg/Fe] and [Si/Fe] as an example in Figure \ref{tub_tag}, tells us that the material stripped earlier from the Sgr progenitor was also less chemically evolved, born from material that experienced fewer Type Ia SNe relative to core collapse SNe, and was therefore either formed earlier in Sgr's history or formed in regions with slower chemical enrichment.  The slight differences between the Mg and Si gradients may then inform us about the detailed nucleosynthetic production of these elements, and how they differ over the star formation history of Sgr.

\section{Discussion}

In this work, we report not only the existence of chemical abundance differences between the Sgr core and the Sgr stream, but along the stream itself.  These abundance variations imply a significant population gradient within the Sgr stream, with the lower metallicity and higher $\alpha$-element abundance populations that were, on average, born from less enriched material than the dominant populations still found in the Sgr dSph core today.

As has been previously suggested \citep{chou2007,keller2010,lm10,carlin2012,shi2012,hyde2015}, the abundance gradients along the stream are thought to be produced by the typically outside-in nature of tidal stripping acting on radial abundance gradients within the Sgr progenitor that are established through its secular chemical evolution.  The particular chemical gradient imprinted on each arm is also dictated by the dynamics of tidal stripping and the differential angular spreading that occurs between the leading and trailing arms and at different phases along each arm itself.  These dynamical variations complicate the direct comparison and interpretation of the two arms' chemical patterns. 

In principle, a more holistic approach to assessing these gradients is possible by estimating the dynamical ages of individual stream stars using the \citetalias{lm10} model of the Sgr stream.  With each stream star timestamped to a specific dynamical stripping age, we can more accurately combine the data from the two tidal arms to reveal and map the significant change in the chemistry of different populations that were pulled from the Sgr progenitor over time (Fig. \ref{tub_tag}).  To the degree that tidal stripping preserves the relative radial distribution of stars in the Sgr progenitor, the abundance gradients we measure with dynamical age should correlate with the initial radial abundance gradients in the Sgr progenitor.  Therefore, we should expect that the Sgr progenitor had an increasing fraction of more metal-poor and $\alpha$-enhanced stars with increasing radius.  This can give us an idea of the magnitude of the chemical differences that might have existed within the progenitor, but, unfortunately, to ultimately reconstruct the actual radial chemical profile of the Sgr progenitor requires knowledge of the original density (stellar {\it and} dark matter) profile of that system.

To simplify this problem, \citetalias{lm10} approached it from the other direction -- i.e., by assuming a density profile for the progenitor, painting the constituent particles with abundances according to a prescription based on the energies of the particles, and then dynamically evolving the system to produce tidal stream abundance gradients.  By varying the chemical abundance prescription with a fixed progenitor density profile (which could also be made a free parameter, but was not), the model stream gradients were constrained to match those observed, but were also traced back to report the requisite radial metallicity profile.  More specifically, to approximately reproduce the metallicity distributions observed in the Sgr stream \citep{chou2007,monaco2007}, \citetalias{lm10} applied a metallicity distribution to the starting satellite configuration in their N-body simulation, which used a Plummer model \citep{plummer1911}, prescribing systematically lower metallicities to the higher energy particles, which would typically populate larger radii and be stripped earlier from the Sgr progenitor.  

With this approach, \citetalias{lm10} found that to produce a 0.6 dex metallicity difference between the present-day Sgr stream and core in their model -- i.e., similar to the largest differences we find between the core and tidal arms here -- required an initial average metallicity variation of $\sim 2.0$ dex from the center to the edge of the Sgr progenitor.  The inferred mean metallicity variation with radius within the original galaxy exceeds that observed along the stream because the tidal stripping of Sgr occurs primarily when the core is at pericenter.  Due to the tidal impulse during pericenter passage, it is not just the outermost stars that are stripped away; instead, these episodes can dredge up stars from deeper in the galaxy's potential well, mixing stars from different orbital radii, blending populations, and producing a shallower gradient in the stream \citepalias{lm10}.

Thus, any observed mean metallicity and $\alpha$-element abundance differences observed along the Sgr streams define the minimum radial variation that existed in the Sgr progenitor.  According to the modeling by \citetalias{lm10}, these variations may have been much larger, such that they far exceed those seen in any dwarf satellite of the MW today \citep[e.g., the $\sim1$ dex metallicity gradient in Sculptor;][]{tolstoy2004}, making Sgr anomalous in this regard.  

However, to place the original Sgr mean radial abundance variation properly in the context of those of other present-day dwarf galaxies, we may need to account for several potentially complicating factors, such as:  (1) The progenitor Sgr system may have been different from other dSphs in the MW halo today in some way.  (2) Other dwarf galaxies may also have had larger radial abundance variations in the past, but, like Sgr, these have also been reduced by tidal stripping, although perhaps to a lesser degree.  (3) Current models of the Sgr tidal disruption \citep{law2005,lm10,lokas2010,penarrubia2010,teppergarcia2018} are incomplete (e.g., they cannot yet account for the stream bifurcation), rudimentary (particularly concerning the details of the star formation and gaseous evolution of the Sgr core), and, in some regards, are even at odds with observations, such as improperly matching the northern portion of the trailing arm \citep{hernitschek2017,sesar2017}; thus, inferences drawn from such models must be considered tentative.

Regarding scenario (1), it is clear that the present Sgr system {\it looks} different than most other MW satellites in several ways.  First, the present Sgr core is among the most massive dSphs, ranking second only  to Fornax.  But even Fornax only exhibits a $\sim 0.7$ dex drop in mean metallcity from center to edge \citep{battaglia2006,leaman2013}.  However, also clearly setting Sgr apart is that it is the one classical (i.e., more massive) satellite that is obviously undergoing major tidal disruption, which is leading to a substantial loss of stars from the core.  If one accounts for the lost stars {\it and} dark matter, the mass of the original Sgr may have far exceeded that of Fornax, perhaps placing the Sgr progenitor in the mass range between that of the SMC and LMC \citep{chou2010,mucciarelli2017,gibbons2017,carlin2018}.

It has been suggested that the Sgr progenitor may have been a disk galaxy that is in the process of being ``tidally stirred'' into a highly stretched dSph morphology  \citep{lokas2010,penarrubia2010}, although the Sgr core does not exhibit evidence of significant rotation \citep{penarrubia2011,frinchaboy2012} that may be expected of such a system.  Nonetheless, if Sgr was initially a disk galaxy, given estimates of its former mass, the Sgr progenitor may well have resembled the disky LMC, which could have further differentiated the Sgr progenitor from present-day classical dSphs.  

On the other hand, even if the Sgr progenitor was a galaxy like the LMC before being tidally stripped and stirred, this may not yet explain the large inferred mean metallicity variation implied across Sgr by modeling.  Even the LMC today does not seem to exhibit as large a radial mean metallicity drop, with only a $\sim 0.5$ dex difference from the LMC center to the $r \sim 10$ kpc extent that has been studied to date \citep{cioni2009,choudhury2016}.  Thus, even compared with the Magellanic satellites, the inferred ~2.0 dex mean metallicity variation across the Sgr progenitor seems extreme.

Alternatively (scenario 2),  perhaps the tidal processes affecting Sgr are not quite so unique.  Like Sgr, the other dwarf satellites may have experienced tidal stripping that removed their least bound, most metal-poor populations, and produced the smaller metallicity variations seen in them today.  Indeed, as an example of this phenomenon we can look at the present Sgr core itself, a system where we know that there has been significant stripping of metal-poor stars, and today exhibits only a $\sim$ 0.2-0.3 dex metallicity range with radius \citep{majewski2013,mucciarelli2017} -- {\it significantly} smaller than the $\sim 0.6$ dex difference observed between the Sgr core and the dynamically old parts of the stream, and not even close to the 2 dex initial radial difference in the progenitor inferred from modeling \citepalias{lm10}.  

While some of the most massive MW dSphs, such as Fornax and Sculptor, seem to have too large of orbits to be significantly affected by tidal stripping \citep{battaglia2015,iorio2019}, there is some support for the notion that other MW dSphs have been affected by tidal stripping \citep{majewski2002,munoz2006,sohn2007,battaglia2012,roderick2015,roderick2016}, and with that stripping preferentially removing older, more metal-poor populations to shape the overall present-day metallicity distribution functions \citep{majewski2002,munoz2006,chou2007,lm10,sales2010,battaglia2012}.  If this is true, then present-day dwarfs may have smaller radial metallicity variations than in the past, because they have also experienced some tidal evolution, albeit not as strongly as Sgr. Indeed, the relatively {\it small} metallicity variation in the Sgr core today may simply reflect that the Sgr orbit is smaller, yielding closer and more frequent pericenter passages that create stronger tidal evolution.  In this scenario, Sgr may well have started out as a more typical dwarf galaxy prior to its currently strong tidal interaction with the MW.

As for scenario (3), we have already identified above several deficiencies in the current Sgr disruption models.  More sophisticated and self-consistent models would better account for the observed gradients in the Sgr stream and enable a more appropriate comparison of the Sgr progenitor with other MW dwarf satellite galaxies.  For example, as noted by \citetalias{lm10}, one of the most obvious inconsistencies in their model \citep[and other models of the Sgr stream; ][]{penarrubia2010,teppergarcia2018}, is that standard N-body models do not self-consistently account for the continuing star formation that the Sgr core experienced over the several billion years that the system tidally evolved within the MW's own changing external potential.

In the \citetalias{lm10} model, particles are assigned ages and metallicities {\it ab initio}, including star particles that should be born after the beginning of the simulation.  While no particles are stripped before they would nominally be born, there are particles that are assigned ages that would require them to be born during the tidal stripping of Sgr.  Not only will the gas in the Sgr progenitor evolve differently than stars \citep{teppergarcia2018} as Sgr orbits the MW, but the populations born during the tidal stripping of Sgr are some of the most bound, metal-rich particles populating the inner radii of the initial model.  Therefore, these young stars (model particles) raise the initial metallicity of the Sgr progenitor core and contribute to a steepening of the metallicity gradient in the initial Sgr model when, in reality, stars of those metallicities would not have been born until several billion years after the tidal stripping of Sgr began.

A more sophisticated and self-consistent modeling of Sgr as an evolving system of stars {\it and} gas would lend considerable insights into the interaction between the star formation, chemical, and dynamical evolution of the system as a whole.  For example, it is known that the Sgr core has experienced a relatively bursty star formation history \citep{siegel2007}, but it is not clear how much of this was induced or modulated by Sgr's interaction with the MW (e.g., compressional shocking of gas at pericenter sparking star formation or the complex effects of ram pressure stripping that facilitate some gas loss, but can compress the remaining gas to produce more star formation), and whether this bursty star formation only occurred in the central-most regions of the Sgr core, or if this bursty star formation was more widespread.  Star formation induced by interaction with the MW that rapidly accelerated the enrichment of the Sgr core could have produced the young, metal-rich populations seen in the Sgr dSph core today, but not contributed to the populations seen in the Sgr stream, which were relatively ``frozen'' for the past several billion years.

As mentioned above, more than one of these various complicating factors may have contributed to explaining the larger inferred radial abundance variation in the Sgr progenitor compared to those observed in more standard dSph satellites of the MW.

\section{Conclusion}

We present an abundance analysis of the Sgr system using data from the largest sample of Sgr stream stars having high-resolution spectra to date.  This stream sample, mostly obtained serendipitously in the course of the APOGEE survey, totals 166 stream members, 63 of which are in the trailing arm and 103 in the leading arm.  We identified these stars as belonging to the Sgr stream by their kinematics, derived via a combination of {\it Gaia} DR2 proper motions, \texttt{StarHorse} spectrophotometric distances \citep{santiago2016,starhorse}, and APOGEE radial velocities.

In particular, we have selected these stream members based on the consistency of their angular momentum with the Sgr core and their kinematical alignment with respect to the Sgr orbital plane.  This kinematical selection quite cleanly identifies Sgr stream members free of most other MW contamination.  We use this sample of Sgr stream stars, together with the 325 Sgr dSph core members from \citet{has17} plus 385 additional core members identified here to measure the metallicity and $\alpha$-element abundance differences between and variations along the Sgr stream and core system.

We find a considerable metallicity difference of $\Delta$[Fe/H] $\sim 0.6$ dex and $\alpha$-element abundance differences of $\Delta$[Mg/Fe] $\sim 0.06$ dex and $\Delta$[Si/Fe] $\sim 0.15$ dex between the Sgr dSph core and the dynamically older, leading arm Sgr stream subsample.  Our 
trailing arm subsample, which is dynamically younger, falls in between the core and leading arm in both metallicity and $\alpha$-element abundances.  These differences indicate that there are metallicity and $\alpha$-element gradients from the Sgr core through each arm of the Sgr stream.  However, we typically find much flatter gradients -- consistent with zero -- when we measure gradients internally, along each of the sampled parts of each tidal arm separately (not including the higher metallicity Sgr dSph core), except for some evidence that there is still a significant $\alpha$-element gradient internally along the leading arm. 

Past modeling has shown that most of the tidal stripping of Sgr occurs in episodes during pericenter passages, and consulting such models, we find that our leading and trailing arm samples each primarily trace material stripped during different pericenter passages, but likely do not individually contain many stars that come from multiple stripping episodes.  Therefore neither arm sample explored here {\it would be expected} to exhibit strong internal gradients with Sgr longitude.

By prescribing dynamical (i.e., stripping) ages from the \citetalias{lm10} model onto our stream sample, we can combine our samples from both arms into a more integrated view, which demonstrates that there are metallicity and $\alpha$-element abundance gradients as a function of dynamical age {\it across} the stream as a whole.  This provides better evidence that there were radial abundance gradients within the Sgr progenitor, because it is expected that the dynamical age of stars map more directly onto their initial orbital radius or total energy within the former Sgr galaxy.

Previous modeling of the tidal evolution of Sgr has found that episodic tidal impulses during pericenter passages will dredge up and mix multiple stellar populations from different radii as they are stripped away, reducing the abundance variations seen when stars are pulled into the tidal stream \citepalias{lm10}.  Conversely, any abundance variations seen in the tidal stream today should betray stronger radial variations in the initial system.

Modeling suggests that the initial radial metallicity variations in the Sgr progenitor might have been very large indeed \citepalias[as much as 2 dex in overall metallicity;][]{lm10}, far exceeding those observed in any present-day MW satellites.  However, we argue that such a large inferred abundance variation compared to other present-day dSphs might be partially explained if the Sgr progenitor had been structurally different than the other systems, e.g., more massive or perhaps morphologically different (e.g., a dwarf spiral, like the LMC).  Moreover, the abundance variations in other dwarfs may have also been reduced (though to a lesser extent) by tidal stripping.  

Further interpretation of the gradients now confirmed to exist along the Sgr stream, would, however, benefit from more sophisticated modeling of the Sgr system, to determine how steep the initial abundance gradients within the Sgr progenitor must have been.  The greatest improvement in current models would be self-consistent treatment of star formation and chemical enrichment as Sgr evolves under the tidal influence of the MW.  This modeling could reveal how much the core of Sgr has evolved since it fell into the MW's potential, and how much mixing occurred during it's episodic stripping.

The presence of primarily more metal-poor Sgr stars in the Sgr stream demonstrates how studies of the present Sgr core will yield skewed metallicity and $\alpha$-element distribution functions compared to those that were actually produced over time in the original Sgr system.  Only by combining the growing data set of high-resolution spectroscopy of Sgr stream stars with samples of the Sgr dSph core that are consistently analyzed can we accurately reconstruct the chemical abundance profile of the Sgr progenitor. We will present such an analysis in \citet{hayes2019inprep}, using multiple elements produced via different nucleosynthetic pathways, to better understand the chemical evolution of the Sgr system.  

\acknowledgements
The authors would like to thank the anonymous referee for their helpful and constructive comments.  This research made use of {\sc topcat} \citep{topcat}.  CRH acknowledges the NSF Graduate Research Fellowship through Grant DGE-1315231.  SRM was funded by NSF grant AST-1616636. S.H. is supported by an NSF Astronomy and Astrophysics Postdoctoral Fellowship under award AST-1801940.  Support for this work was provided by NASA through Hubble Fellowship grant \#51386.01 awarded to RLB by the Space Telescope Science Institute, which is operated by the Association of  Universities for Research in Astronomy, Inc., for NASA, under contract NAS 5-26555.

CAP acknowledges funding from the Spanish MINECO grant AYA2017-86389-P.  G.B. gratefully acknowledges financial support through the grant (AEI/FEDER, UE) AYA2017-89076-P and the MCIU Ram\'on y Cajal Fellowship RYC-2012-11537, as well as by the Ministerio de Ciencia, Innovaci\'on y Universidades (MCIU), through the State Budget and by the Consejer\'\i a de Econom\'\i a, Industria, Comercio y Conocimiento of the Canary Islands Autonomous Community, through the Regional Budget.  TCB acknowledges partial support from grant PHY 14-30152; Physics Frontier Center/JINA Center for the Evolution of the Elements (JINA-CEE), awarded by the US National Science Foundation, and from the Leverhulme Trust (UK), during his visiting professorship at the University Of Hull, when this paper was finished.  PMF acknowledges support for this research from the National Science Foundation (AST-1715662).  DAGH acknowledges support from the State Research Agency (AEI) of the Spanish Ministry of Science, Innovation and Universities (MCIU) and the European Regional Development Fund (FEDER) under grant AYA2017-88254-P. SzM has been supported by the Premium Postdoctoral Research Program and J{\'a}nos Bolyai Research Scholarship of the Hungarian Academy of Sciences, by the Hungarian NKFI Grants K-119517 and GINOP-2.3.2-15-2016-00003 of the Hungarian National Research, Development and Innovation Office. R.R.M. acknowledges partial support from project BASAL AFB-$170002$ as well as FONDECYT project N$^{\circ}1170364$.

Funding for the Sloan Digital Sky Survey IV has been provided by the Alfred P. Sloan Foundation, the U.S. Department of Energy Office of Science, and the Participating Institutions. SDSS-IV acknowledges support and resources from the Center for High-Performance Computing at the University of Utah. The SDSS web site is www.sdss.org.

SDSS-IV is managed by the Astrophysical Research Consortium for the Participating Institutions of the SDSS Collaboration including the Brazilian Participation Group, the Carnegie Institution for Science, Carnegie Mellon University, the Chilean Participation Group, the French Participation Group, Harvard-Smithsonian Center for Astrophysics, Instituto de Astrof\'isica de Canarias, The Johns Hopkins University, Kavli Institute for the Physics and Mathematics of the Universe (IPMU) / University of Tokyo, the Korean Participation Group, Lawrence Berkeley National Laboratory, Leibniz Institut f\"ur Astrophysik Potsdam (AIP),  Max-Planck-Institut f\"ur Astronomie (MPIA Heidelberg), Max-Planck-Institut f\"ur Astrophysik (MPA Garching), Max-Planck-Institut f\"ur Extraterrestrische Physik (MPE), National Astronomical Observatories of China, New Mexico State University, New York University, University of Notre Dame, Observat\'ario Nacional / MCTI, The Ohio State University, Pennsylvania State University, Shanghai Astronomical Observatory, United Kingdom Participation Group,Universidad Nacional Aut\'onoma de M\'exico, University of Arizona, University of Colorado Boulder, University of Oxford, University of Portsmouth, University of Utah, University of Virginia, University of Washington, University of Wisconsin, Vanderbilt University, and Yale University.

\end{document}